\documentclass[preprint,12pt,authoryear]{elsarticle}

\usepackage{amssymb}
\usepackage{amsmath}
\usepackage{booktabs}

\usepackage{bm}

\usepackage{xcolor}
\colorlet{revision}{black}
\newcommand{\rev}[1]{{\color{revision}#1}}

\usepackage{tikz}
\usetikzlibrary{arrows.meta,calc,fit,backgrounds}
\usepackage{xurl}
\usepackage[hidelinks]{hyperref}
\providecommand{\doi}[1]{\href{https://doi.org/#1}{\nolinkurl{#1}}}

\graphicspath{{figures/}}

\newcommand{\RE}{Re_{\lambda_g,i}}  % fixed case identifier
\newcommand{\Relam}{Re_{\lambda_g}}  % Taylor scale Re
\newcommand{\ustar}{\overline{u_1u_1}^{\,*}}
\newcommand{\estar}{\varepsilon^{*}}
\newcommand{\tstar}{t^{*}}

\begin{document}

\hypersetup{
  pdftitle={Construction of a Deployable Closure Model for the Transport of Turbulence Dissipation Rate with Hard-Constrained Physics-Informed Machine  Learning},
  pdfauthor={Tamerlan Hajiyev, Emre Sefer, Ivan Otić, Ismail Ari, {\"O}zg{\"u}r Ertun\c{c}}
}

\begin{frontmatter}



\title{Construction of a Deployable Closure \rev{Model for the Transport of Turbulence Dissipation Rate with Hard-Constrained Physics-Informed Machine Learning}} %% Article title


\author[cs]{Tamerlan Hajiyev}
\ead{tamerlan.hajiyev@ozu.edu.tr}
\author[ai]{Emre Sefer}
\ead{emre.sefer@ozyegin.edu.tr}
\author[kit]{Ivan Otić}
\ead{ivan.otic@kit.edu}
\author[cs]{Ismail Ari}
\ead{ismail.ari@ozyegin.edu.tr}
\author[me]{{\"O}zg{\"u}r Ertun\c{c}\corref{cor1}}
\ead{ozgur.ertunc@ozyegin.edu.tr}
\cortext[cor1]{Corresponding author}

%% Author affiliations
\affiliation[cs]{organization={Department of Computer Science, {\"O}zye{\u{g}}in University},
            city={Istanbul},
            country={T\"urkiye}}
\affiliation[ai]{organization={Department of Artificial Intelligence and Data Engineering, {\"O}zye{\u{g}}in University},
            city={Istanbul},
            country={T\"urkiye}}
\affiliation[kit]{organization={Karlsruhe Institute of Technology},
            city={Karlsruhe},
            country={Germany}}
\affiliation[me]{organization={Fluid Dynamics and Spray Laboratory, Department of Mechanical Engineering, {\"O}zye{\u{g}}in University},
            city={Istanbul},
            country={T\"urkiye}}

%% Abstract
\begin{abstract}
We present a \rev{hard-constrained physics-informed machine learning (PIML) method for
constructing a deployable closure model for the transport of turbulence
dissipation rate}.
The framework is built from modules, each performing a single task.
The closure module represents the closure with a small network, while the
state module represents the flow state during training. The closure network
is coupled to the state network
through the governing equation residuals\rev{. This coupling allows us to learn} the closure
without using reference closure values as training \rev{targets}.
The coordinate module learns monotone input coordinates that
help both networks resolve sharp changes. Within the network
architecture, the boundary condition (BC) module imposes boundary
values exactly using the Theory of Functional Connections, and the
realizability module keeps the closure realizable. \rev{We study decaying homogeneous
isotropic turbulence (HIT) and learn $\psi$, the closure function governing
dissipation rate decay, from trajectories obtained in experiments and direct
numerical simulations.} Here, the state module is
a Deep Operator Network (DeepONet), the BC module imposes the initial
state and the closure's low and high Reynolds number limits, and the
realizability module ensures growth of the Taylor microscale. The framework reduces the root mean square of the closure
variation across ten random seeds by about 95\% relative to the unconstrained model over the training
and held-out range. The deployed 18-parameter closure network requires only
the Taylor scale Reynolds number $\Relam$ and can be exported as a standalone
C function.
%% End of abstract
\end{abstract}

\begin{graphicalabstract}
% Figure 1: modular closure construction (requires tikz with the arrows.meta,
% calc, fit and backgrounds libraries).
% The layout follows the available width: the two parts sit side by side on a
% single-column page or across both columns, and are stacked in one column.
% Box text is set at 8 pt and frame titles at 12 pt bold in every format.
\begingroup%
\newif\ifarchnarrow%
\ifdim\linewidth<11cm\relax\archnarrowtrue\else\archnarrowfalse\fi%
\definecolor{archCoord}{HTML}{6A3D9A}%
\definecolor{archState}{HTML}{1F78B4}%
\definecolor{archClosure}{HTML}{D95F02}%
\definecolor{archBC}{HTML}{1B9E77}%
\definecolor{archReal}{HTML}{CB181D}%
\newcommand{\archfont}{\fontsize{8}{9.6}\selectfont}%
\newcommand{\archtag}[2]{\textcolor{#1}{\fontsize{7}{8.4}\selectfont\bfseries #2}}%
\newcommand{\archu}{\widehat{\overline{u_1u_1}^{\,*}}}%
\newcommand{\arche}{\widehat{\varepsilon^{*}}}%
\tikzset{
  archmod/.style={draw=#1, fill=#1!7, line width=0.7pt, rounded corners=2pt,
    align=center, inner sep=2.5pt, font=\archfont},
  archsub/.style={draw=#1, fill=white, line width=0.5pt, rounded corners=1.5pt,
    align=center, inner sep=2pt, font=\archfont},
  archbox/.style={draw=black!45, fill=black!3, line width=0.5pt,
    rounded corners=2pt, align=center, inner sep=2.5pt, font=\archfont},
  archarrow/.style={-{Stealth[length=4pt,width=3pt]}, line width=0.6pt,
    draw=black!70},
  archpart/.style={draw=black!45, dashed, line width=0.6pt, rounded corners=4pt},
  archtitle/.style={font=\fontsize{12}{14.4}\selectfont\bfseries, text=black!75,
    inner sep=1pt, align=center},
}%
\sbox0{%
\begin{tikzpicture}[x=1cm, y=1cm, font=\archfont]
  % State part, identical in both layouts.
  \node[archsub=archState, text width=1.8cm] (bu) at (6.40,-0.95)
    {branch $\mathbf b_{u_1u_1}$};
  \node[archsub=archState, text width=1.8cm] (be) at (6.40,-1.80)
    {branch $\mathbf b_\epsilon$};
  \node[archsub=archState, text width=1.8cm] (tr) at (6.40,-3.10)
    {trunk $\boldsymbol\phi$};
  \node[anchor=south, inner sep=1pt, align=center, text=archState,
    font=\fontsize{7}{8.4}\selectfont\bfseries] (stag)
    at ($(bu.north)+(0,0.08)$) {State module\\(DeepONet)};
  \begin{scope}[on background layer]
    \node[archmod=archState, fit=(stag)(bu)(be)(tr), inner sep=4pt] (state) {};
  \end{scope}
  % Inputs and their coordinates, level with the branches and the trunk.
  \coordinate (bmid) at ($(bu)!0.5!(be)$);
  \node[archmod=archCoord, text width=3.5cm] (qr) at (3.05,0 |- bmid)
    {\archtag{archCoord}{Coordinate module}\\ $q_R(Re_{\lambda_g,i})$};
  \node[archmod=archCoord, text width=3.5cm] (qt) at (3.05,0 |- tr)
    {\archtag{archCoord}{Time coordinate module}\\ $q_t(t^*)$};
  \node (rei) at (0.30,0 |- qr) {$Re_{\lambda_g,i}$};
  \node (tin) at (0.30,0 |- qt) {$t^*$};
  \node[archmod=archBC, text width=3.65cm] (bcs) at (10.18,-2.05)
    {\archtag{archBC}{Boundary condition (BC)}\\ \archtag{archBC}{module using TFC}\\[1pt]
     $\archu=1+\mathbf b_{u_1u_1}^{\mathsf T}\Delta\boldsymbol\phi$\\[1pt]
     $\arche=1+\mathbf b_{\epsilon}^{\mathsf T}\Delta\boldsymbol\phi$\\[1pt]
     $\Delta\boldsymbol\phi=\boldsymbol\phi(t^*)-\boldsymbol\phi(0)$};
  \draw[archarrow] (rei) -- (qr);
  \draw[archarrow] (tin) -- (qt);
  \draw[archarrow] (qr.east) -- ++(0.25,0) |- (bu.west);
  \draw[archarrow] (qr.east) -- ++(0.25,0) |- (be.west);
  \draw[archarrow] (qt) -- (tr);
  \draw[archarrow] (state.east |- bcs.west) -- (bcs.west);
  \node[archpart, fit=(qr)(qt)(state)(bcs), inner sep=8pt] (spart) {};
  \node[archtitle, anchor=south west] at (spart.north west)
    {State network};

  % Coupling and closure part: a column beside the state part, or two rows
  % of two modules below it.
  \ifarchnarrow
    \coordinate (cpl) at ($(3.4,0 |- spart.south)+(0,-0.75)$);
    \coordinate (cq) at ($(cpl)+(0,-1.8)$);
    \coordinate (cg) at ($(cpl)+(5.2,-1.8)$);
    \coordinate (cb) at ($(cpl)+(0,-3.4)$);
    \coordinate (cs) at ($(cpl)+(5.2,-3.4)$);
    \def\archcw{3.65cm}
    \def\archcplanchor{center}
  \else
    % The closure column starts near the top of the state network frame; the
    % training box ends level with the bottom of the closure frame.
    \coordinate (cpl) at ($(16.05,0 |- spart.north)+(0,-0.37)$);
    \coordinate (cq) at ($(cpl)+(0,-2.00)$);
    \coordinate (cg) at ($(cpl)+(0,-3.35)$);
    \coordinate (cb) at ($(cpl)+(0,-5.00)$);
    \coordinate (cs) at ($(cpl)+(0,-6.64)$);
    \def\archcw{3.65cm}
    \def\archcplanchor{north}
  \fi
  \node[archbox, anchor=\archcplanchor] (cplnode) at (cpl)
    {$\widehat{Re}_{\lambda_g}=\dfrac{Re_{\lambda_g,i}\,\archu}{\sqrt{\arche}}$};
  \node[archmod=archCoord, text width=\archcw] (qpsi) at (cq)
    {\archtag{archCoord}{Coordinate module}\\ $q_\psi(\widehat{Re}_{\lambda_g})$};
  \node[archmod=archClosure, text width=\archcw] (mlp) at (cg)
    {\archtag{archClosure}{Closure module (MLP)}\\ $g(q_\psi)$};
  \node[archmod=archBC, text width=\archcw] (bcc) at (cb)
    {\archtag{archBC}{Boundary condition (BC)}\\ \archtag{archBC}{module using TFC}\\ $\psi(0)=1.4$\\ $\psi(Re_{\lambda_g,\infty})=11/6$};
  \node[archmod=archReal, text width=\archcw] (real) at (cs)
    {\archtag{archReal}{Realizability module}\\ $\psi=1+\sigma(h(\widehat{Re}_{\lambda_g}))>1$};
  \draw[archarrow] (cplnode) -- (qpsi);
  \draw[archarrow] (qpsi) -- (mlp);
  \ifarchnarrow
    \draw[archarrow] (bcs.south) |- (cplnode.east);
    \draw[archarrow] (mlp.south) -- ++(0,-0.22) -| (bcc.north);
    \draw[archarrow] (bcc) -- (real);
    \node[archpart, fit=(qpsi)(mlp)(bcc)(real), inner sep=8pt] (cpart) {};
    \node[archtitle, anchor=south east] at (cpart.north east)
      {Deployable\\ closure network};
    \node (psi) at ($(real.south |- cpart.south)+(0,-0.35)$)
      {$\psi(\widehat{Re}_{\lambda_g})$};
    \draw[archarrow] (real) -- (psi);
  \else
    \draw[archarrow] (bcs.north) |- (cplnode.west);
    \draw[archarrow] (mlp) -- (bcc);
    \draw[archarrow] (bcc) -- (real);
    \node[archpart, fit=(qpsi)(mlp)(bcc)(real), inner sep=8pt] (cpart) {};
    \node[archtitle, anchor=south] at ($(cplnode.north)+(0,0.1)$)
      {Deployable\\ closure network};
  \fi

  % Training.
  \ifarchnarrow
    \node[archbox, text width=9.7cm, anchor=north west] (train)
      at ($(cpart.south west |- psi.south)+(0,-0.35)$)
      {\archtag{black!80}{Training}\\[1pt]
       $R_{u_1u_1}=\dfrac{Re_{\lambda_g,i}}{10}\dfrac{d\archu}{dt^*}+\arche$\\[2pt]
       $R_\epsilon=\dfrac{Re_{\lambda_g,i}}{10}\dfrac{d\arche}{dt^*}
         +\psi\,\dfrac{(\arche)^2}{\archu}$\\[2pt]
       $\mathcal L=\mathcal L_{\rm data}+\lambda_{\rm phys}
         \left(\langle R_{u_1u_1}^2\rangle+\langle R_\epsilon^2\rangle\right)$};
    \draw[archarrow] (psi.south) -- (psi.south |- train.north);
    \coordinate (inmid) at ($(rei)!0.5!(tin)$);
    \draw[archarrow] (spart.west |- inmid) -- ($(rei.west |- inmid)+(-0.25,0)$)
      |- (train.west);
  \else
    \node[archbox, text width=10.9cm, anchor=south west] (train)
      at (spart.west |- cpart.south)
      {\archtag{black!80}{Training}\\[1pt]
       $R_{u_1u_1}=\dfrac{Re_{\lambda_g,i}}{10}\dfrac{d\archu}{dt^*}+\arche,
       \qquad
       R_\epsilon=\dfrac{Re_{\lambda_g,i}}{10}\dfrac{d\arche}{dt^*}
         +\psi\,\dfrac{(\arche)^2}{\archu}$\\[2pt]
       $\mathcal L=\mathcal L_{\rm data}+\lambda_{\rm phys}
         \left(\langle R_{u_1u_1}^2\rangle+\langle R_\epsilon^2\rangle\right)$};
    \draw[archarrow] (bcs.south) -- (bcs.south |- train.north);
    \draw[archarrow] (real.west) -- (real.west -| train.east)
      node[midway, above, inner sep=2pt] {$\psi(\widehat{Re}_{\lambda_g})$};
  \fi
\end{tikzpicture}}%
\ifdim\wd0>\linewidth\relax\resizebox{\linewidth}{!}{\usebox0}\else\usebox0\fi%
\endgroup%

\end{graphicalabstract}

\begin{highlights}
\item The physics-informed machine learning framework enables building closures of unclosed terms in the transport \rev{equation} of turbulent dissipation
\item The developed method makes use of experimental and direct numerical simulation data within $2<Re_{\lambda_g}<1499$ along with the theoretical limits at $Re_{\lambda_g}\rightarrow0$ and $Re_{\lambda_g}\rightarrow \infty$  to model the decay of homogeneous isotropic turbulence
\item Boundary conditions and realizability hold exactly by construction \rev{for homogeneous isotropic turbulence}
\item The framework reduces RMS closure variation across seeds by 95\%
\item A DeepONet state module enables learning without reference closure data
\item The closure network takes only $\Relam$ as input; together with its coordinate map, it has 18 learned parameters
\end{highlights}

%% Keywords
\begin{keyword}
turbulence closure \sep physics informed \sep DeepONet \sep deployable closure \sep hard constraints \sep learned input coordinates \sep homogeneous isotropic turbulence



\end{keyword}

\end{frontmatter}


%% main textrecm
%%

\section{Introduction}
\label{sec:intro}
%% Labels are used to cross-reference an item using \ref command.

Turbulence closures model the unclosed terms in the averaged Navier--Stokes
equations and in the associated Reynolds stress and dissipation rate
transport equations. Classical models combine theoretical analysis with experimental
and numerical data, while accounting for invariance, realizability, and known
limiting behavior \citep{launder1975, pope1975, lumley1978, LAUNDER1989, Speziale_Sarkar_Gatski_1991, jovanovic2003}.
Data-driven models allow more flexible closure relations, but they must still
satisfy these physical requirements.

These requirements can be incorporated into closure modeling using
physics-informed neural networks (PINNs)
\citep{raissi2019pinn,karniadakis2021}.
They commonly enforce constraints through loss penalties
\citep{wang2021pideeponet,mcconkey2025realisability}, but these do not guarantee
exact satisfaction. Architectural constraints can instead enforce specific
properties, such as tensor invariance and total energy conservation, by
construction \citep{ling2016,vangastelen2024}.
\rev{For their Reynolds stress models,} \citet{zhang2026bluff} jointly
\rev{trained} state and closure networks using \rev{reference stress values}
and \rev{deployed} the learned models in an independent solver.
\rev{They used a prescribed or learned length scale to model dissipation
in the turbulent kinetic energy equation without solving a separate
dissipation rate equation.}
\rev{They enforced realizability by construction and imposed boundary
conditions through loss penalties.} Boundary conditions can be
enforced exactly using trial functions that combine a term satisfying them
with a free function that cannot alter them \citep{lagaris1998}. The Theory
of Functional Connections (TFC) provides a general construction of such
functions, and Deep TFC uses a neural network as the free function
\citep{mortari2017tfc,leake2020deeptfc}.

\rev{We study decaying homogeneous isotropic turbulence (HIT), for which the
dissipation rate equation takes the form $d\varepsilon/dt=-\psi\,\varepsilon^2/k$, where $k$ is
the turbulent kinetic energy and $\varepsilon$ its dissipation rate.
The term $-\psi\,\varepsilon^2/k$ models $P_\varepsilon^4-\gamma$,
the difference between production by vortex stretching and viscous
destruction \citep{jovanovic2003}.
The standard $k$--$\varepsilon$ model uses a constant $C_{\varepsilon2}=1.92$
in place of $\psi$ \citep{pope2000}, while \citet{Speziale_Sarkar_Gatski_1991}
used $C_{\varepsilon2}=1.83$.
\citet{Hanjalic_Launder_1976} modeled how this coefficient varies with Reynolds number
in decaying grid turbulence.
\citet{mansour1994} evaluated the terms in the dissipation rate equation
directly from DNS, while \citet{jovanovic2003} proposed an explicit model
for $\psi$ based on the ratio of the Taylor and integral length scales.}
\citet{gunseren} \rev{inferred the decay function in their
nondimensional dissipation rate equation} from \rev{decaying} HIT trajectories,
without \rev{supplying reference closure values} during training\rev{. They adapted} the
\rev{trained} network to new cases through transfer learning, but did not enforce
physical constraints by construction or demonstrate independent solver deployment.

Operator learning has also been used for closure modeling. Deep Operator
Networks (DeepONets) use a trunk network to learn basis functions of the
evaluation coordinates and a branch network to learn input dependent
coefficients. Their inner product approximates the operator output
\citep{lu2021deeponet}. \citet{taassob2024} combine PINNs and DeepONet to
reconstruct flow fields and estimate closure terms in turbulent combustion
using \rev{flow measurements} and \rev{estimates of} chemical source terms as
training data, but do not demonstrate independent solver deployment.

Given these studies, a general framework would bring these ideas together. It
would learn deployable turbulence closures from state trajectories without reference
closure values as training data, enforce boundary conditions and realizability
by construction, and allow additional physical constraints to be incorporated
for more complex flows. \rev{The PIML-based modeling framework has the potential to reduce the time required to develop turbulence closures. We have selected the decaying homogeneous isotropic turbulence case to put the results in context.} The framework should also address
sensitivity to input representation and training initialization. 

To address these requirements, we develop a modular framework for constructing
a compact \rev{closure for the dissipation rate equation in decaying HIT}.
The PIML algorithm learns the closure relation jointly with the state representation
through the governing equation residuals, without prescribing an analytical
closure model or supplying reference closure values as training data.
The framework enforces boundary conditions and physical constraints exactly, while learned
monotone input coordinates reduce sensitivity to training initialization.
We prescribe the
Batchelor limit $\psi=1.4$ for the final period of decay \citep{batchelor1948}, the Saffman
high Reynolds number limit $\psi=11/6$ \citep{saffman1967}, and the condition
$\psi>1$ associated with growth of the Taylor microscale. We assess trajectory
prediction and closure consistency across random seeds, and evaluate the
learned closure in an independent numerical integrator. The closure is
exported with its coordinate and constraint modules as a standalone C
function, without retaining the DeepONet state network.

\section{Decaying HIT model and data}
\label{sec:problem}

This section introduces the decaying HIT problem and the data used
to demonstrate the framework of Section~\ref{sec:method}.

\subsection{Physical setting and nondimensional variables}
\label{subsec:nondimensional}

Decaying homogeneous isotropic turbulence is described using the
single-component velocity variance and the dissipation rate. For the \rev{wind
tunnel measurements of} grid-generated decaying HIT,
\rev{we use} Taylor's hypothesis and the local mean velocity
\rev{to convert downstream distance to time}. Measured streamwise and transverse
variances are combined into the pseudo-isotropic single-component variance
\begin{equation}
  \overline{u_1u_1}^{\mathrm{eff}}
    = \tfrac13\bigl(\overline{u_1u_1}+2\,\overline{u_2u_2}\bigr).
\end{equation}
We use the single-component RMS velocity convention of \citet{pope2000}.
The initial Taylor scale Reynolds number and characteristic time are
\begin{equation}
  \RE = \frac{\sqrt{15}\;\overline{u_1u_1}^{\mathrm{eff}}_{\,i}}
             {\sqrt{\nu\,\varepsilon_i}},
  \qquad
  t_c = \sqrt{15\nu/\varepsilon_i},
  \label{eq:relgi}
\end{equation}
where the subscript $i$ denotes the initial state of each trajectory.
Thus, $\tstar=t/t_c$, $\ustar=\overline{u_1u_1}^{\mathrm{eff}}/
\overline{u_1u_1}^{\mathrm{eff}}_{\,i}$, and
$\estar=\varepsilon/\varepsilon_i$, with $\ustar(0)=\estar(0)=1$ by
construction. The Taylor scale Reynolds number along the
decay is
\begin{equation}
  \Relam(\tstar) = \RE\,\frac{\ustar(\tstar)}{\sqrt{\estar(\tstar)}} .
  \label{eq:relam}
\end{equation}
Here, $\RE$ is the fixed initial Reynolds number that identifies each case,
whereas $\Relam$ evolves during decay and serves as the closure input.

The formulation of \citet{jovanovic2003} uses the full fluctuating magnitude
$q=\sqrt{3}\,u_{\mathrm{rms}}$ under local isotropy. Its Reynolds
number is therefore larger by $\sqrt{3}$:
\begin{equation}
  \Relam = Re_{\lambda_g}^{\,q}/\sqrt3 .
  \label{eq:conv}
\end{equation}

\subsection{Governing transport equations}
\label{subsec:governing}

We use the two-equation decay system of \citet{gunseren}. Using the notation above,
the system reads
%% align or eqnarray environments can be used for multi line equations.
%% & is used to mark alignment points in equations.
%% \\ is used to end a row in a multiline equation.
\begin{align}
  \frac{d\ustar}{d\tstar} &= -\frac{10}{\RE}\,\estar,
    \label{eq:ode_u}\\[2pt]
  \frac{d\estar}{d\tstar} &= -\psi(\Relam)\,\frac{10}{\RE}\,
     \frac{(\estar)^2}{\ustar},
    \label{eq:ode_e}
\end{align}
where $\psi\rev{(\Relam)}$ is the dimensionless \rev{coefficient governing the decay of the dissipation rate}.
Equation~\eqref{eq:ode_u} governs the normalized single-component velocity
variance, while Eq.~\eqref{eq:ode_e} governs the normalized dissipation rate
and contains the closure.

%% Use \subsubsection, \paragraph, \subparagraph commands to
%% start 3rd, 4th and 5th level sections.
%% Refer following link for more details.
%% https://en.wikibooks.org/wiki/LaTeX/Document_Structure#Sectioning_commands

\subsection{Features and limiting values of closure model}
\label{subsec:closure-hyp}

Following \citet{jovanovic2003}, we assume that the closure depends on the
state only through the Taylor scale Reynolds number. Within each dataset,
we represent the closure by a single function $\psi(\Relam)>1$. We prescribe the two limiting values
\begin{equation}
  \psi(0)=1.4, \qquad \psi(Re_{\lambda_g,\infty})=\frac{11}{6},
  \qquad Re_{\lambda_g,\infty}=10^6.
  \label{eq:psi_limits}
\end{equation}
The lower value follows \rev{from the decay law for} the final period
\citep{batchelor1948}, and the
upper value is the Saffman limit \citep{saffman1967}. The upper computational
limit $Re_{\lambda_g,\infty}=10^6$ represents $\Relam\to\infty$.

\citet{jovanovic2003} define
\begin{equation}
 \begin{aligned}
  W(\Relam)
  &=0.626\Bigl[-0.049\sqrt{3}\,\Relam\\
  &\qquad+\frac12\sqrt{0.009604\,
    \bigl(\sqrt{3}\,\Relam\bigr)^2+10.208}\Bigr].
 \end{aligned}
 \label{eq:jovanovic_weight}
\end{equation}
\begin{equation}
 \psi_{\mathrm J}(\Relam)=1.8-0.4W(\Relam).
 \label{eq:jovanovic_closure}
\end{equation}
The model is included only as a reference curve and tends to $1.8$ at very high
Reynolds numbers.

In the framework (Section~\ref{sec:method}), the BC module imposes the two values in
Eq.~\eqref{eq:psi_limits} exactly, and the realizability module enforces $\psi>1$.
Thus, $\psi(0)=1.4$ is exact, while $1.4$ is not imposed as a lower bound away from that endpoint.
No interior monotonicity, derivative, or curvature condition is prescribed. The TFC expression acts
over the full computational interval, while the free network determines the
remaining interior variation.

\subsection{Data sources and trajectory reconstruction}
\label{subsec:data}
We use data from experiments and Direct Numerical Simulations (DNS) for training and testing.  Table~\ref{tab:datasets} summarizes the data sources and the three datasets. Dataset~1, the first six rows in Table~\ref{tab:datasets}, is mainly experimental, and Dataset~2 combines DNS \citep{yoffe2018} with the
Comte-Bellot--Corrsin (CBC) \citep{cbc1971} and Variable Density Turbulence Tunnel (VDTT) \citep{sinhuber2015} cases, and Dataset~3 combines both.

\begin{table}[htbp]
\centering
\resizebox{\linewidth}{!}{%
\begin{tabular}{llcccccc}
\toprule
& & & & & \multicolumn{3}{c}{Dataset} \\
\cmidrule(lr){6-8}
Source & Type & $\Relam$ range & Train & Held out & 1 & 2 & 3 \\
\midrule
Grid generated HIT, Ertun\c{c} (2007)      & measured  & $13.1$--$35.5$   & 5 & 1 & \checkmark &            & \checkmark \\
HIT chamber, Co\c{s}ar et al.\ (2026) & measured & $278.3$--$450.9$ & 2 & 0 & \checkmark &            & \checkmark \\
Burattini et al.\ (2006)            & DNS & $19.2$--$49.0$   & 1 & 1 & \checkmark &            & \checkmark \\
Lavoie et al.\ (2007)               & measured & $37.3$--$43.0$   & 0 & 1 & \checkmark &            & \checkmark \\
Comte-Bellot and Corrsin (1971)     & measured & $58.9$--$71.6$   & 1 & 0 & \checkmark & \checkmark & \checkmark \\
Sinhuber et al.\ (2015)             & measured & $93.4$--$1498.7$ & 5 & 3 & \checkmark & \checkmark & \checkmark \\
Yoffe and McComb (2018)             & DNS       & $2.03$--$44.1$   & 4 & 3 &            & \checkmark & \checkmark \\
\midrule
\multicolumn{5}{l}{Training trajectories} & 14 & 10 & 18 \\
\multicolumn{5}{l}{Held-out trajectories} &  6 &  6 &  9 \\
\bottomrule
\end{tabular}}
\caption{Data sources and the three datasets. Each $\Relam$ range spans the
trajectories of that source.}
\label{tab:datasets}
\end{table}

% \subsubsection{Power-law reconstruction}
% \label{par:reconstruction}
The grid-generated HIT measurements and DNS of \citet{yoffe2018} provide
trajectories. \rev{For all other cases, we use the} decay exponents \rev{to
reconstruct power-law}
trajectories with the same number of data points. 

For a case with exponent $n$, let $t_{\mathrm{rel}}$ be the time elapsed since the
initial state and $t_0$ the time from the virtual origin of the power law to
the initial state. The decay ratio $r=1+t_{\mathrm{rel}}/t_0$ then measures
time from the virtual origin, with $r=1$ at the initial state, and
\begin{equation}
  \ustar = r^{-n}, \qquad \estar = r^{-(n+1)}, \qquad
  \tstar = \frac{t_{\mathrm{rel}}}{t_c}
         = (r-1)\,\frac{n\,\RE}{10},
  \label{eq:reconstruct}
\end{equation}
which satisfies Eqs.~\eqref{eq:ode_u}--\eqref{eq:ode_e} with the constant value
$\psi=(n+1)/n$ \citep{mansour1994}. The Reynolds number is
$\Relam=\RE r^{(1-n)/2}$. We use $r\in[1,5]$, which covers the main decay
phase and excludes the late tail, where the signal approaches the noise level.
For the HIT chamber cases, $t_0=0.1$~s is the reference time of the experimental
power-law fits.

We convert six grid-generated HIT trajectories measured by \citet{ertunc2007} to nondimensional form using the definitions above, with $\varepsilon$ obtained from the decay of the turbulent kinetic energy. These data were taken as part of the PhD thesis but were never published in their raw form,  and are made available with the paper.

\citet{cosar2026ilass} \rev{used planar particle image velocimetry (PIV) to obtain}
the HIT chamber measurements in a $30\times30\times30$ cm propeller-driven
chamber, also known as a French washing machine. Turbulence
was first forced to a statistically stationary state with the rotating propellers according to a given protocol. The propellers were then
stopped, and the decay of turbulence was recorded at 10 Hz. Ten independent realizations
were synchronized at the stopping time. Decay exponents were fitted to the pooled measurements. We used the reference fits for the cases at 3000 and 4000 rpm for training.

Decay exponents are taken from the lattice Boltzmann simulations of
\citet{burattini2006} and the experiments of \citet{lavoie2007},
\citet{cbc1971}, and \citet{sinhuber2015}. One Burattini case is used for
training and one is held out. The representative Lavoie Rd35-WSA case is held
out, and the CBC case is used for training. The VDTT study of
\citet{sinhuber2015} contains 99 decay curves acquired at 36 Reynolds numbers
with three hot-wire probes. The eight VDTT cases used here represent
nonoverlapping logarithmic Reynolds number bins of the measurements
digitized from their figures.
Five are used for training, and the three cases above $\RE=800$ are held out.

We select seven decaying DNS cases from the dataset of
\citet{yoffe2018}. With the Taylor microscale defined by
$\lambda_g^2=15\nu\,\overline{u_1u_1}/\varepsilon$, power-law decay gives
$\lambda_g^2=10\nu(t-t_v)/n$, which grows linearly from the virtual origin
$t_v$ \citep{yoffe2018}. For each case, the initial state is therefore set at
the start of an approximately linear segment of $\lambda_g^2(t)$. The
energy decay exponent fitted over the selected region is used only to
distribute the sampled records over a common decay interval; the DNS state
values themselves are taken directly from the simulations. Eighteen actual
records after the initial state are selected over $r\in(1,5]$. Four cases are
used for training and three are held out.

% \paragraph{Wind-tunnel trajectories}
% \label{par:winddata}

% \paragraph{HIT chamber PIV measurements}
% \label{par:restdata}

% \paragraph{Published cases}

\section{\texorpdfstring{Closure construction framework}{Closure construction framework}}
\label{sec:method}

\begin{figure}[htbp]
\centering
\input{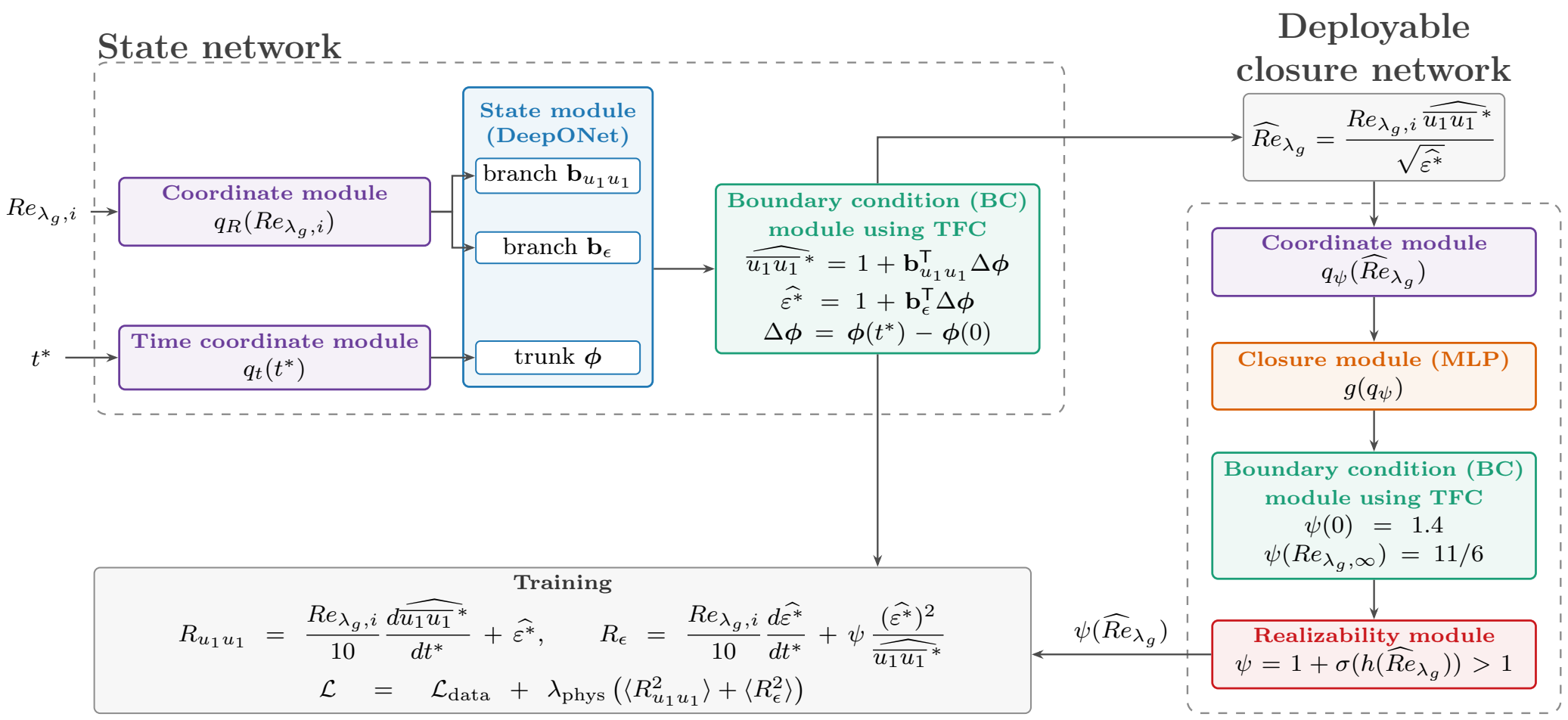}
\caption{Closure construction framework for decaying HIT. The frame labeled "State network"
contains the state module with its coordinate and BC modules, used during
training, and the frame labeled "Deployable closure network" contains the closure
module with its coordinate, BC, and realizability modules, retained for
deployment.
Colors mark the module type. The two frames are coupled through $\Relam$ and the
dissipation residual. MLP: multilayer perceptron; TFC: Theory of Functional
Connections.}
\label{fig:architecture}
\end{figure}

The framework learns a deployable closure from trajectories without requiring reference closure values as training data. In decaying HIT, this closure is
$\psi(\Relam)$ in Eq.~\eqref{eq:ode_e}. It is built from modules that each
perform a single task. The modules form two networks that are trained together
(Fig.~\ref{fig:architecture}): the state network represents the decaying state during training, and the closure network represents $\psi$. The two networks are
coupled only through the residuals of Eqs.~\eqref{eq:ode_u}--\eqref{eq:ode_e},
evaluated at collocation points spaced by $\Delta\tstar\approx0.1$. After
training, only the closure is retained.

The constraints imposed by the modules hold by construction. Each module can be
replaced independently\rev{. New modules can enforce physical
constraints,} such as invariance and realizability, in more complex turbulent flows.

\subsection{\texorpdfstring{Closure module}{Closure module}}
\label{subsec:closure-module}

The closure module represents the unknown closure as a function of the
chosen closure inputs. For \rev{decaying HIT}, it represents $\psi$ by a
multilayer perceptron (MLP) $g$
with two hidden layers of width 2 and tanh activations\rev{. The MLP} has 12
trainable parameters. Its only input is the coordinate $q_\psi(\Relam)$
(Section~\ref{subsec:coordinates}), so the closure depends on the state only
through $\Relam$.

\subsection{\texorpdfstring{State module}{State module}}
\label{subsec:state-surrogate}

The closure is not supervised; it enters only through the governing
equations, here the dissipation equation. To learn it through the
governing equation residuals, the state of every training case must
be available as a differentiable function of time. The state module
provides this representation during training.

The state module uses a DeepONet \rev{to represent each} output as the
inner product of a branch vector that depends on the case and a trunk
vector that depends on time. Each branch receives the coordinate $q_R$ of the
initial Reynolds number $\RE$,  and the trunk
receives the coordinate $q_t$ of $\tstar$ (Section~\ref{subsec:coordinates}):
\begin{equation}
  \mathbf b_{u_1u_1}=\mathbf b_{u_1u_1}(q_R(\RE)), \qquad
  \mathbf b_\epsilon=\mathbf b_\epsilon(q_R(\RE)), \qquad
  \boldsymbol\phi=\boldsymbol\phi(q_t(\tstar)).
\end{equation}
The components of $\boldsymbol\phi$ act as temporal basis functions, and
$\mathbf b_{u_1u_1}$ and $\mathbf b_\epsilon$ are their case-dependent
coefficients. Because Eqs.~\eqref{eq:ode_u}--\eqref{eq:ode_e} are coupled, we assume that both states can be represented by the same basis functions, so one trunk is shared by the two branches, while separate branches retain the
distinct dependence of the two states on $\RE$. The embedding dimension, the
number of basis functions, is 10. Each branch has two hidden layers of width
15, and the trunk has four hidden layers of width 15. All three networks use
SiLU activations. The state module contains 1760 trainable parameters.

\subsection{\texorpdfstring{Coordinate module}{Coordinate module}}
\label{subsec:coordinates}

The coordinate module learns a monotone map of each network input, so that
training decides where to place resolution. A small network that sees its
inputs through a fixed scaling may not have enough resolution where sharp changes occur. For decaying HIT, the inputs span several decades, and the
coordinate ablations in Section~\ref{subsec:results-fixed-coordinates} confirm
that, with fixed coordinates, the learned closure depends on the chosen
scaling. Each map has two stages: a reference coordinate $z\in[0,1]$ sets the
overall scaling, and a Bernstein map refines the resolution within it.

For the case and time inputs, neither a linear nor a logarithmic scaling
is known to be appropriate in advance, so the reference coordinate is a
learnable blend of the two,
\begin{equation}
\begin{split}
  z_R={}&(1-a_R)\frac{\RE-Re_{\lambda_g,i}^{\min}}{Re_{\lambda_g,i}^{\max}-Re_{\lambda_g,i}^{\min}} \\
      &+a_R\frac{\log(1+\RE)-\log(1+Re_{\lambda_g,i}^{\min})}
                {\log(1+Re_{\lambda_g,i}^{\max})-\log(1+Re_{\lambda_g,i}^{\min})},
\end{split}
  \label{eq:branch-coordinate}
\end{equation}
\begin{equation}
  z_t=(1-a_t)\frac{\tstar}{t^*_{\rm max}}
      +a_t\frac{\log(1+\tstar)}{\log(1+t^*_{\rm max})},
  \label{eq:time-coordinate}
\end{equation}
where $Re_{\lambda_g,i}^{\min}$ and $Re_{\lambda_g,i}^{\max}$ are the
smallest and largest initial Reynolds numbers of the training trajectories,
$t^*_{\rm max}$ is the largest observation time of the training
trajectories, and $0<a_R<1$ and $0<a_t<1$.

The closure input covers $[0,Re_{\lambda_g,\infty}]$, required by the BC module, while the training data cover only
$[Re_{\lambda_g}^{\min},Re_{\lambda_g}^{\max}]$. Since $Re_{\lambda_g,\infty}=10^6$ and
$Re_{\lambda_g}^{\max}$ is below $10^3$, the data range occupies less than 0.1\% of a
linear coordinate, so a small network would spend almost all of its resolution
on the tail. The reference coordinate \rev{assigns a share of $[0,1]$ to each of}
three intervals\rev{. These cover the range} below the data, the data range
and the tail. With
$x=\Relam/Re_{\lambda_g,\infty}$, the knots are
\begin{equation}
\begin{split}
  (x_0,x_1,x_2,x_3)&=\left(0,\frac{Re_{\lambda_g}^{\min}}{Re_{\lambda_g,\infty}},
  \frac{Re_{\lambda_g}^{\max}}{Re_{\lambda_g,\infty}},1\right), \\
  (y_0,y_1,y_2,y_3)&=(0,m_0,m_0+m_1,1),
\end{split}
\end{equation}
where the reference masses $m_k$ give the share of $[0,1]$ assigned to
each interval. On interval $[x_k,x_{k+1}]$, let $w_k=x_{k+1}-x_k$,
$h_k=y_{k+1}-y_k$, $\delta_k=h_k/w_k$, and $\xi=(x-x_k)/w_k$. The reference
coordinate is the monotone rational-quadratic spline of
\citet{durkan2019neural},
\begin{equation}
  C(x)=y_k+h_k
  \frac{\delta_k\xi^2+d_k\xi(1-\xi)}
       {\delta_k+(d_{k+1}+d_k-2\delta_k)\xi(1-\xi)},
  \label{eq:closure-reference-coordinate}
\end{equation}
with endpoint derivatives $d_0=\delta_0$ and $d_3=\delta_2$ and internal
derivatives
\begin{equation}
  d_k=
  \frac{2\delta_{k-1}\delta_k}{\delta_{k-1}+\delta_k}
  \exp(\beta_k), \qquad k=1,2,
  \label{eq:closure-coordinate-derivatives}
\end{equation}
where $\beta_k$ is trainable and initialized at zero. With $\eta_2=0$, the
reference masses are
\begin{equation}
  m_k=0.02+0.94\frac{\exp(\eta_k)}
  {\sum_{j=0}^{2}\exp(\eta_j)}, \qquad k=0,1,2,
\end{equation}
initialized at 5\%, 90\%, and 5\%, so most of the initial resolution lies
on the data range.

The two stages are complementary: for the closure, the reference coordinate concentrates
most of the initial resolution on the data range, but some regions within it
still require higher resolution than others. Each reference coordinate is then
refined by a Bernstein polynomial of degree $m$, which adjusts the
resolution further,
\begin{equation}
  B_m(z)=\sum_{j=0}^{m}\alpha_j
       \binom{m}{j}z^j(1-z)^{m-j},
  \label{eq:bernstein-map}
\end{equation}
with $\alpha_0=0$, $\alpha_m=1$, and increasing coefficients. Increasing
coefficients make the map strictly increasing \citep{ramasinghe2022bernstein},
so input order is preserved, the fixed end coefficients keep the endpoints in
place, and the Bernstein basis is numerically stable under evaluation
\citep{farouki1996stability}. Every coefficient increment is at least 0.01,
and a fixed reference logit leaves $m-1$ independent increment parameters.
The mapped inputs are $q_R=2B_{14}(z_R)-1$, $q_t=2B_{14}(z_t)-1$, and
$q_\psi=2B_3(C(x))-1$. The maps $q_R$ and $q_t$ each learn a blend weight and
13 Bernstein increments, and $q_\psi$ learns two reference masses, two internal
derivatives, and two Bernstein increments, six parameters in total.

\subsection{\texorpdfstring{BC module}{BC module}}
\label{subsec:constrained-closure}

Physical models generally have prescribed boundary conditions, and they should be modeled faithfully. For decaying HIT, every normalized trajectory starts from
$(\ustar,\estar)=(1,1)$, and the closure must take the endpoint values in
Eq.~\eqref{eq:psi_limits}. A penalty would only approximate these values, so
the BC module imposes them exactly with the Theory of Functional Connections
\citep{mortari2017tfc,leake2020deeptfc}. A TFC expression adds \rev{a correction} to a free
function\rev{, replacing} its values at the boundary points with the prescribed
values. The \rev{resulting expression} satisfies the boundary \rev{conditions}
for \rev{any choice of the} free function.

For the state module, the free function is the DeepONet output
$\mathbf b^{\mathsf T}\boldsymbol\phi$, and the only boundary point is
$\tstar=0$:
\begin{align}
  \Delta\boldsymbol\phi(\tstar)
    &=\boldsymbol\phi(q_t(\tstar))-\boldsymbol\phi(q_t(0)), \\
  \widehat{\ustar}
    &=1+\mathbf b_{u_1u_1}^{\mathsf T}\Delta\boldsymbol\phi, \\
  \widehat{\estar}
    &=1+\mathbf b_\epsilon^{\mathsf T}\Delta\boldsymbol\phi,
  \label{eq:state-tfc}
\end{align}
so $\widehat{\ustar}(0)=\widehat{\estar}(0)=1$. A constant added to
$\boldsymbol\phi$ cancels in $\Delta\boldsymbol\phi$, so the trunk carries no
output bias.

For the closure, the free function is $g(q_\psi(\Relam))$ and both ends of
$[0,Re_{\lambda_g,\infty}]$ are boundary points. With
$s=\Relam/Re_{\lambda_g,\infty}$, which increases linearly from 0 to 1 between
the two endpoints, the TFC expression is
\begin{equation}
  h(\Relam)=g(q_\psi(\Relam))-(1-s)\,g(-1)-s\,g(1)+(1-s)h_0+s\,h_1,
  \label{eq:closure-tfc}
\end{equation}
since the coordinate module maps the endpoints to $q_\psi=-1$ and
$q_\psi=1$. At $s=0$ the free terms cancel and $h=h_0$; at $s=1$, $h=h_1$. The
coefficients of the three evaluations of $g$ sum to zero, so an output bias of
$g$ cancels and is omitted. The boundary values $h_0$ and $h_1$ are given in
the latent space of the realizability module
(Section~\ref{subsec:realizability}).

\subsection{\texorpdfstring{Realizability module}{Realizability module}}
\label{subsec:realizability}

A free network can return an unrealizable closure, in particular outside
the training range. We impose $\psi>1$,
which ensures growth of the Taylor microscale since the governing equations give $d\lambda_g^2/dt=10\nu(\psi-1)$. The
realizability module enforces this with the softplus function
$\sigma(x)=\log(1+e^x)$, which is smooth and positive for every $x$. The closure is
\begin{equation}
  \psi(\Relam)=1+\sigma(h(\Relam)).
  \label{eq:positive-closure}
\end{equation}
The prescribed endpoint values are
mapped into the latent space as
\begin{equation}
  h_0=\sigma^{-1}(0.4), \qquad
  h_1=\sigma^{-1}(5/6), \qquad
  \sigma^{-1}(y)=\log(e^y-1),
\end{equation}
so $\psi(0)=1.4$ and $\psi(Re_{\lambda_g,\infty})=11/6$ hold exactly and
$\psi>1$ everywhere. No monotonicity is imposed on $\psi$; its interior is left
to the closure module.

\subsection{Joint training}
\label{subsec:joint-training}

The two networks are trained together: the data loss fits the state module
to the trajectories, and the residuals couple the closure to it. All
trajectories are normalized by their initial state, so every case enters on
the same scale. With $N_{\rm case}$ training trajectories and $N_t=18$ samples
in each, the data loss is
\begin{equation}
\begin{split}
  \mathcal L_{\rm data}
  ={}&\frac{1}{N_{\rm case}}\sum_{c=1}^{N_{\rm case}}
   \frac{1}{N_t}\sum_{i=1}^{N_t}
   \Bigl[
   \left(\widehat{\ustar}_{i,c}-{\ustar}^{\rm data}_{i,c}\right)^2 \\
   &\qquad
   +\left(\widehat{\estar}_{i,c}-{\estar}^{\rm data}_{i,c}\right)^2
   \Bigr],
\end{split}
  \label{eq:data-loss}
\end{equation}
where hats denote network outputs. The nested averages give every
trajectory equal weight, and all trajectories are evaluated together in one
full-batch calculation. The residuals are evaluated at collocation points
placed uniformly in $\tstar$ from 0 to the last observation time of each
training trajectory, with spacing $\Delta\tstar\approx0.1$ and at most 2000
points per trajectory:
\begin{align}
  R_{u_1u_1} &=\frac{\RE}{10}\frac{d\widehat{\ustar}}{d\tstar}
        +\widehat{\estar}, \\
  R_\epsilon &=\frac{\RE}{10}\frac{d\widehat{\estar}}{d\tstar}
        +\psi(\widehat{\Relam})
         \frac{(\widehat{\estar})^2}{\widehat{\ustar}},
  \label{eq:training-residuals}
\end{align}
where $\widehat{\Relam}=\RE\,\widehat{\ustar}/\sqrt{\widehat{\estar}}$ is
the Reynolds number computed from the network outputs. Isolating the time
derivatives would introduce an explicit factor $10/\RE$ in the sensitivity to
$\psi$ and reduce the influence of higher Reynolds number trajectories; the
chosen form avoids this without an additional loss weight.
\begin{equation}
  \mathcal L=\mathcal L_{\rm data}
    +\lambda_{\rm phys}
     \left(\langle R_{u_1u_1}^2\rangle+\langle R_\epsilon^2\rangle\right),
  \label{eq:joint-loss}
\end{equation}
where each bracket averages over the collocation points of one trajectory and
then equally over the training trajectories. Numerical clamps prevent division
by zero and are inactive for all solutions across the three datasets.

Training starts with 500 epochs of the data loss alone, so the state module first represents the trajectories before the residuals involve the untrained
closure. The physics weight then increases linearly from zero to one over 500
epochs, which avoids an abrupt change of the objective. The coordinate maps are
frozen during both phases and are learned once the full objective is active.
Training uses 30,000 epochs in total. One Adam optimizer uses separate
parameter groups: network learning rates follow a cosine schedule from
$10^{-3}$ to $10^{-5}$, and coordinate learning rates from $5\times10^{-4}$ to
$5\times10^{-6}$. The combined gradient norm is clipped at 1.0\rev{. We select the checkpoint
with the lowest value of the full training objective}. Held-out trajectories
are used only for evaluation.

\subsection{\texorpdfstring{Deployable closure network}{Extraction and deployment}}
\label{subsec:closure-extraction}

The state module and its coordinates $q_R$ and $q_t$ 
are used only during training. The deployed closure network
consists of the coordinate $q_\psi$, the closure module, and the closure parts of
the BC and realizability modules (Fig.~\ref{fig:architecture}). It accepts $\Relam$ and returns the closure
required by Eq.~\eqref{eq:ode_e}, with no separate dependence on time, the
trajectory label, or the initial Reynolds number.

\rev{For forward predictions, we} integrate Eqs.~\eqref{eq:ode_u}--\eqref{eq:ode_e}
from $(\ustar,\estar)=(1,1)$ using fourth-order Runge--Kutta with step
$\Delta\tstar=0.1$\rev{. We recompute} $\Relam$ at every stage. No DeepONet
prediction enters this integration, so it tests the extracted closure
independently. For standalone use, the trained parameters are written into a
scalar function in the C programming language. The training-only state module
and its coordinates contain 1760 network and 28 coordinate parameters;
the deployed closure network retains 12 network and 6 coordinate parameters.

\section{Experiments and results}
\label{sec:experiments}

Dataset~1 is used to test the modules one at a time: the constraints, the
closure capacity, and the coordinates. The selected construction is then
applied without modification to Datasets~2 and~3, and the three resulting
closures are compared in forward integration.

\subsection{Common protocol}
\label{subsec:common-protocol}

All runs use the state architecture in Section~\ref{subsec:state-surrogate}
and the objective and 30,000-epoch schedule in
Section~\ref{subsec:joint-training}. Each configuration is trained with the ten
seeds $\{42,1,2,3,4,5,6,7,8,9\}$. Held-out trajectories enter neither training
nor checkpoint selection. The ablation labels list the coordinates $q_R$,
$q_t$, and $q_\psi$: ``lin'' and ``log'' denote fixed affine and min--max
scaled logarithmic coordinates, ``raw'' denotes no input normalization, and
``learned'' denotes the coordinate module (Table~\ref{tab:ablations}). The selected closures are also integrated forward without the state module
(Section~\ref{subsec:results-forward}). Reproducibility is measured by the
variation across seeds. With the seed mean
\begin{equation}
  \overline{\psi}_{\rm seed}(R_j)
  =\frac{1}{N_s}\sum_{s=1}^{N_s}\psi_s(R_j),
  \label{eq:seed-mean}
\end{equation}
the RMS seed variation over a stated Reynolds number interval
$\mathcal I$ is
\begin{equation}
  D_\psi(\mathcal I)=\left[
    \frac{1}{N_R}\sum_{j=1}^{N_R}
    \frac{1}{N_s-1}\sum_{s=1}^{N_s}
    \left(\psi_s(R_j)-\overline{\psi}_{\rm seed}(R_j)\right)^2
  \right]^{1/2},
  \label{eq:seed-dispersion}
\end{equation}
evaluated on $N_R=20{,}001$ uniform points; all ten seeds are included.

\begin{table}[htbp]
\centering
\fontsize{10}{12}\selectfont
\setlength{\tabcolsep}{3pt}
\renewcommand{\arraystretch}{1.2}
\begin{tabular}{@{}llccclr@{}}
\toprule
ID & Constraints & $q_R$ & $q_t$ & $q_\psi$ & MLP & $p$ \\
\midrule
A & none & lin & log & lin & 1-2-2-1 & 13 \\
B & endpoints & lin & log & lin & 1-2-2-1 & 12 \\
C & endpoints $+$ $\psi>1$ & lin & log & lin & 1-1-1-1 & 5 \\
D & endpoints $+$ $\psi>1$ & lin & log & lin & 1-2-2-1 & 12 \\
E & endpoints $+$ $\psi>1$ & lin & log & lin & 1-3-3-1 & 21 \\
F & endpoints $+$ $\psi>1$ & lin & log & log & 1-2-2-1 & 12 \\
G & endpoints $+$ $\psi>1$ & log & log & lin & 1-2-2-1 & 12 \\
H & endpoints $+$ $\psi>1$ & raw & raw & raw & 1-2-2-1 & 12 \\
I & endpoints $+$ $\psi>1$ & learned & learned & learned & 1-1-1-1 & 11 \\
\textbf{J} & \textbf{endpoints $+$ $\psi>1$} & \textbf{learned} & \textbf{learned} & \textbf{learned} & \textbf{1-2-2-1} & \textbf{18} \\
K & endpoints $+$ $\psi>1$ & learned & learned & learned & 1-3-3-1 & 27 \\
\bottomrule
\end{tabular}
\caption{Dataset~1 configurations. IDs A--K identify the configurations in
Table~\ref{tab:ladder-summary}. MLP gives the network widths, and $p$ is the
number of deployed learned parameters, including the closure coordinate but
excluding the training-only coordinates $q_R$ and $q_t$. The learned closure
coordinate contributes six parameters. The selected configuration J is bold.}
\label{tab:ablations}
\end{table}

\subsection{Effect of the closure constraints}
\label{subsec:results-constraints}

We first test the BC and realizability modules. The same closure module
is trained in three forms: without constraints (13 parameters, including an
output bias), with the BC module, and with the BC and realizability modules
(12 parameters each). The state module keeps its BC module in all three.

\begin{figure}[htbp]
\centering
\includegraphics[width=\linewidth]{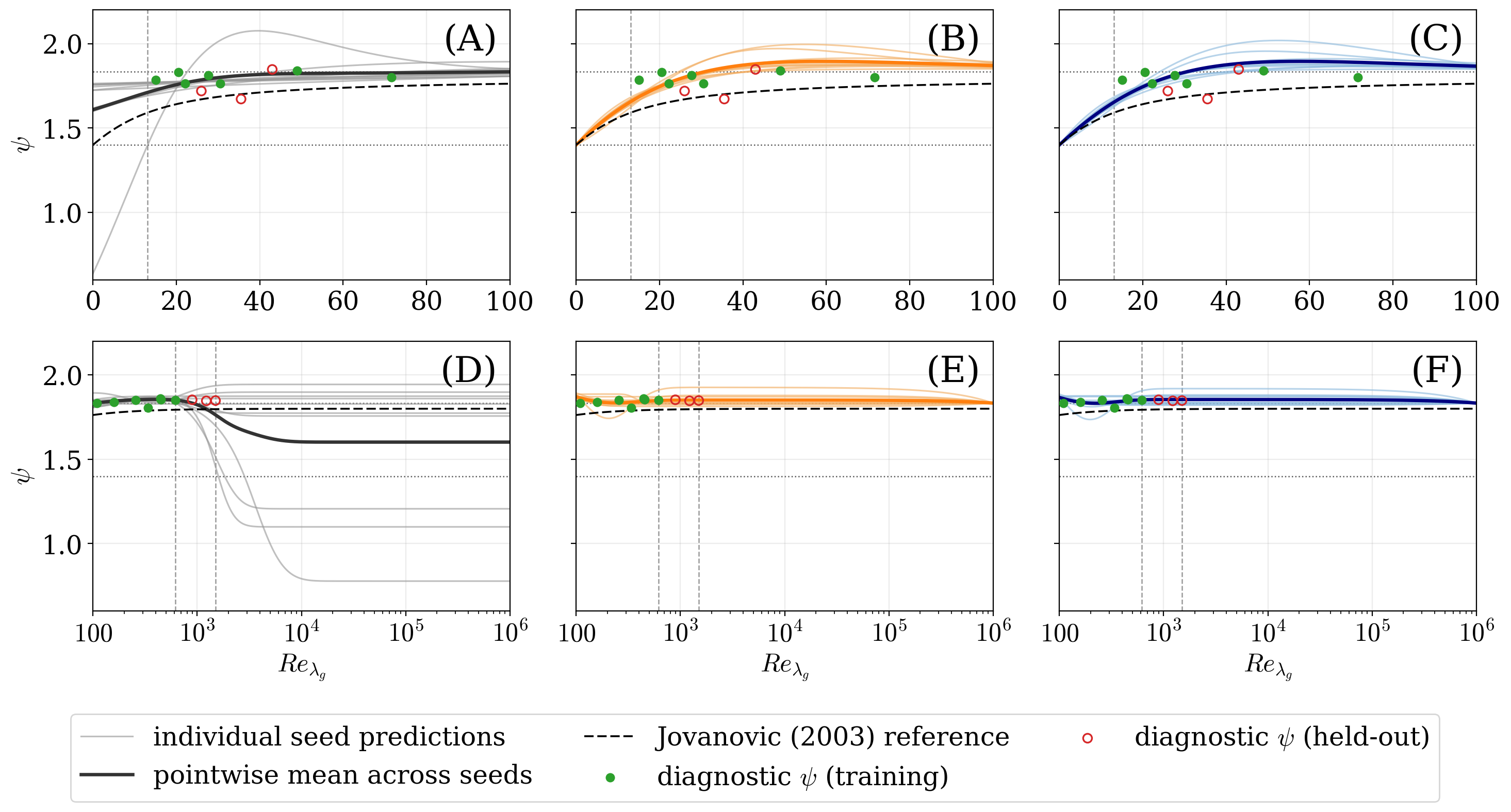}
\caption{Dataset 1 constraint ladder using fixed linear branch, logarithmic
trunk, and linear closure coordinates. (A,D) Unconstrained closure; (B,E)
endpoint TFC; (C,F) endpoint TFC with $\psi>1$. Panels A--C show
$0\leq\Relam\leq100$, and panels D--F show
$100\leq\Relam\leq10^6$. Thin curves are individual seeds and thick curves
are their pointwise means. Symbols show closure values inferred from
trajectory fits for reference; they are not used in training.}
\label{fig:psi-ladder}
\end{figure}

Without constraints, the endpoint values vary strongly between seeds;
$\psi(0)$, for example, ranges from 0.64 to 1.76 (Fig.~\ref{fig:psi-ladder}).
The BC module fixes the endpoints at 1.4 and 11/6 for every seed, and the
realizability module guarantees $\psi>1$ while leaving the interior practically
unchanged. The interior remains free: both constrained closures show a seed-dependent hump near $\Relam\approx58$. To test
whether this hump is physical or a consequence of the representation, we next
vary the closure capacity and the input coordinates.

\subsection{Closure capacity and fixed input coordinates}
\label{subsec:results-fixed-coordinates}

We first test whether more capacity removes the hump, using two hidden layers of width 1, 2, and 3 (5, 12,
and 21 parameters) with the constraints and fixed coordinates unchanged.

\begin{figure}[htbp]
\centering
\includegraphics[width=\linewidth]{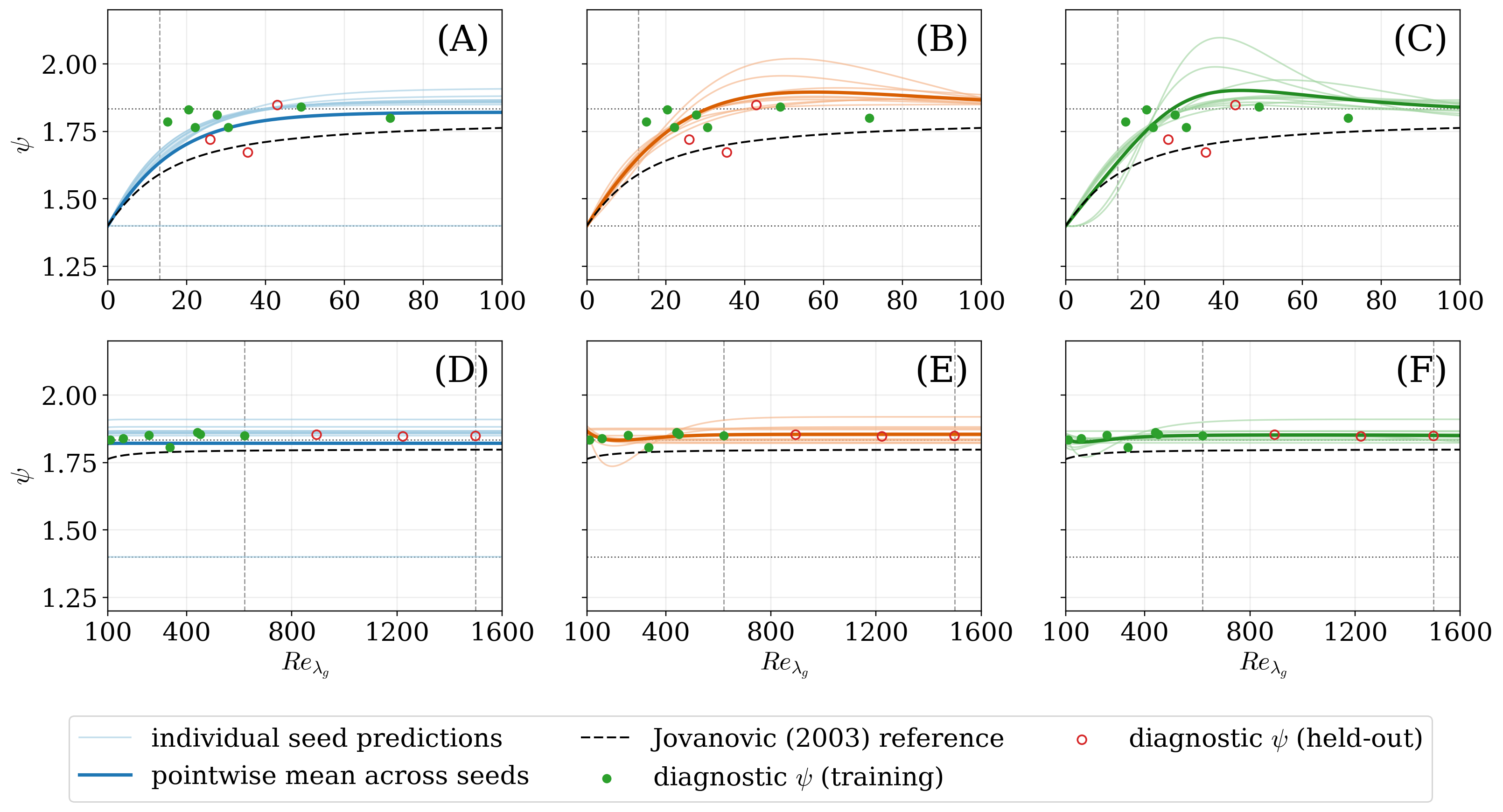}
\caption{Dataset~1 closure capacity comparison for endpoint TFC with
$\psi>1$ using the same fixed coordinates. (A,D)
5-parameter; (B,E) 12-parameter; and (C,F) 21-parameter closure networks.
Panels A--C show $0\leq\Relam\leq100$, and panels D--F show
$100\leq\Relam\leq1600$.}
\label{fig:psi-capacity}
\end{figure}

\rev{For one seed, the} 5-parameter closure remains near $\psi=1.4$
(Fig.~\ref{fig:psi-capacity}). More capacity removes this failure, but the hump becomes more pronounced as
the number of parameters increases. We therefore fix the closure at 12 parameters and change one
prescribed coordinate at a time, with a fourth case that removes all
normalization.

\begin{figure}[htbp]
\centering
\includegraphics[width=\linewidth]{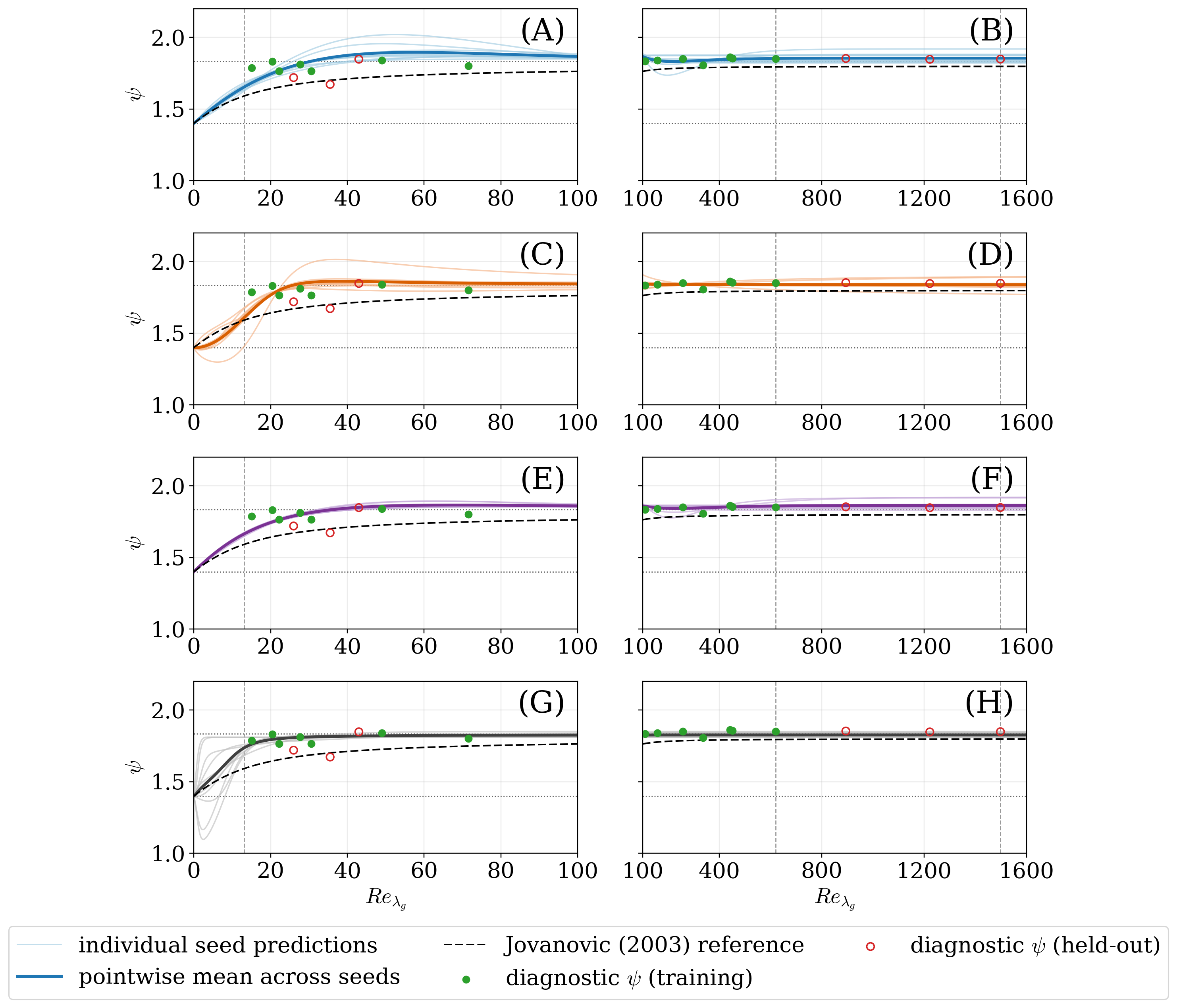}
\caption{Sensitivity to the prescribed coordinates for the 12-parameter
closure. (A,B) Linear branch--linear closure; (C,D) linear branch--log
closure; (E,F) log branch--linear closure; (G,H) no normalization. The left
column shows $0\leq\Relam\leq100$, and the right column shows
$100\leq\Relam\leq1600$.}
\label{fig:psi-fixed-coordinates}
\end{figure}

The unnormalized case reaches the lowest training loss. For four
\rev{seeds, the learned closure falls} below 1.4 at Reynolds numbers below
the data \rev{range}
(Fig.~\ref{fig:psi-fixed-coordinates}). A lower loss therefore does not make a
prescribed coordinate neutral: the learned closure depends on the chosen
representation.

\subsection{Jointly learned coordinates}
\label{subsec:results-joint-coordinates}

More capacity amplifies the hump instead of removing it, which points to a
resolution problem rather than a capacity problem: a fixed coordinate cannot
place resolution where the closure changes quickly. We therefore use the
coordinate module (Section~\ref{subsec:coordinates}), which learns $q_R$,
$q_t$, and $q_\psi$ jointly with the networks. With closure widths of 1, 2, and
3, the deployed closure networks have 11, 18, and 27 parameters, including the six
parameters of $q_\psi$.

\begin{figure}[htbp]
\centering
\includegraphics[width=\linewidth]{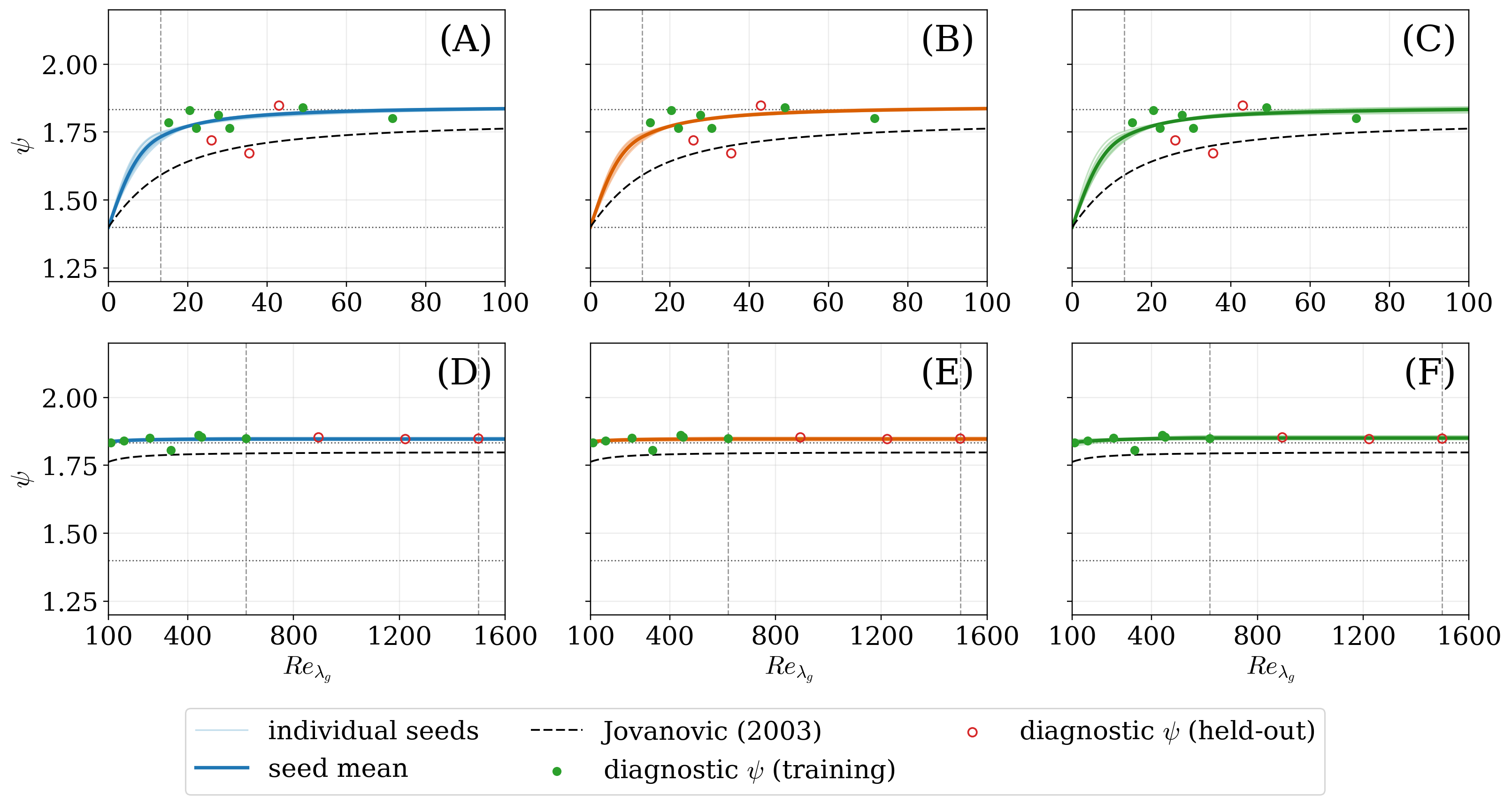}
\caption{Jointly learned coordinates at three closure network capacities.
(A,D) 11 parameters; (B,E) 18 parameters; (C,F) 27 parameters. Panel ranges as in Fig.~\ref{fig:psi-capacity}.}
\label{fig:psi-joint-coordinates}
\end{figure}

All ten seeds now recover nearly the same interior at every capacity, and the
hump disappears (Fig.~\ref{fig:psi-joint-coordinates}). Over the training
range, the RMS \rev{variation across seeds} drops to 0.003--0.004,
compared with 0.014--0.029 for fixed
coordinates, excluding the 5-parameter closure. Of the three
capacities, the 18-parameter closure gives the lowest variation in every
Reynolds number interval, and 27 parameters change the mean closure only slightly, so
the 18-parameter closure is selected.

\subsection{Summary of the Dataset 1 analysis}
\label{subsec:results-ladder}

Table~\ref{tab:ladder-summary} summarizes the Dataset~1 configurations
defined in Table~\ref{tab:ablations}. The closure values implied by the
reported or fitted decay exponents
\rev{serve only as diagnostics.} For seed $s$, $\Delta_{\psi,{\rm data}}$ is the mean
absolute deviation \rev{of the learned closure from these values, given by}
\begin{equation}
  \Delta_{\psi,{\rm data}}^{(s)}=
  \frac{1}{N_c}\sum_{i=1}^{N_c}
  \left|\psi_s(R_i)-\psi_{{\rm data},i}\right|.
  \label{eq:seedwise-discrepancy}
\end{equation}

\begin{table}[p]
\centering
\fontsize{10}{12}\selectfont
\setlength{\tabcolsep}{3pt}
\renewcommand{\arraystretch}{1.15}
\begin{tabular}{@{}lrrrrl@{}}
\toprule
\multicolumn{6}{@{}l}{\textbf{A. Closure accuracy and observed behavior}} \\
\addlinespace[3pt]
& \multicolumn{2}{c}{Training (14 cases)} & \multicolumn{2}{c}{Held out (6 cases)} & \\
\cmidrule(lr){2-3}\cmidrule(lr){4-5}
ID & Mean $\Delta$ & Mean $s_\psi$ & Mean $\Delta$ & Mean $s_\psi$ & Remarks \\
\midrule
A & $0.0326\pm0.0223$ & 0.0359 & $0.0864\pm0.0440$ & 0.0908 & excursions \\
B & $0.0429\pm0.0138$ & 0.0266 & $0.0631\pm0.0288$ & 0.0369 & hump \\
C & $0.0757\pm0.1219$ & 0.1353 & $0.0813\pm0.1114$ & 0.1399 & outlier \\
D & $0.0429\pm0.0143$ & 0.0278 & $0.0645\pm0.0277$ & 0.0373 & hump \\
E & $0.0434\pm0.0195$ & 0.0296 & $0.0715\pm0.0345$ & 0.0467 & hump \\
F & $0.0395\pm0.0210$ & 0.0261 & $0.0687\pm0.0300$ & 0.0396 & dip--hump \\
G & $0.0330\pm0.0055$ & 0.0129 & $0.0503\pm0.0149$ & 0.0203 & hump \\
H & $0.0263\pm0.0031$ & 0.0148 & $0.0552\pm0.0095$ & 0.0124 & dips \\
I & $0.0210\pm0.0009$ & 0.0035 & $0.0419\pm0.0012$ & 0.0038 & reproducible \\
\textbf{J} & $\bm{0.0210\pm0.0006}$ & \textbf{0.0030} & $\bm{0.0416\pm0.0012}$ & \textbf{0.0030} & \textbf{reproducible} \\
K & $0.0208\pm0.0008$ & 0.0043 & $0.0421\pm0.0015$ & 0.0043 & reproducible \\
\bottomrule
\end{tabular}
\par\vspace{16pt}
\begin{tabular}{@{}lrrrrrr@{}}
\toprule
\multicolumn{7}{@{}l}{\textbf{B. Seed reproducibility and training loss}} \\
\addlinespace[3pt]
& \multicolumn{4}{c}{$D_\psi$ by Reynolds number interval} & \multicolumn{2}{c}{Best training loss} \\
\cmidrule(lr){2-5}\cmidrule(lr){6-7}
ID & $0$--$10^6$ & $0$--$13.1$ & $13.1$--$620$ & $620$--$10^6$ & Mean & $s$ \\
\midrule
A & 0.4130 & 0.2411 & 0.0268 & 0.4131 & $2.41\times10^{-4}$ & $2.08\times10^{-4}$ \\
B & 0.0192 & 0.0165 & 0.0281 & 0.0192 & $2.19\times10^{-4}$ & $2.16\times10^{-4}$ \\
C & 0.0871 & 0.0526 & 0.1474 & 0.0870 & $4.30\times10^{-4}$ & $6.56\times10^{-4}$ \\
D & 0.0182 & 0.0164 & 0.0287 & 0.0182 & $2.23\times10^{-4}$ & $2.09\times10^{-4}$ \\
E & 0.0191 & 0.0473 & 0.0247 & 0.0191 & $1.93\times10^{-4}$ & $1.46\times10^{-4}$ \\
F & 0.0404 & 0.0637 & 0.0205 & 0.0404 & $1.37\times10^{-4}$ & $1.05\times10^{-4}$ \\
G & 0.0173 & 0.0074 & 0.0171 & 0.0173 & $7.31\times10^{-5}$ & $1.60\times10^{-5}$ \\
H & 0.0081 & 0.1609 & 0.0140 & 0.0081 & $4.47\times10^{-5}$ & $2.06\times10^{-5}$ \\
I & 0.0022 & 0.0177 & 0.0033 & 0.0022 & $4.75\times10^{-5}$ & $7.65\times10^{-6}$ \\
\textbf{J} & \textbf{0.0020} & \textbf{0.0150} & \textbf{0.0030} & \textbf{0.0020} & $\bm{4.67\times10^{-5}}$ & $\bm{8.22\times10^{-6}}$ \\
K & 0.0030 & 0.0187 & 0.0044 & 0.0030 & $4.64\times10^{-5}$ & $1.04\times10^{-5}$ \\
\bottomrule
\end{tabular}
\caption{Dataset~1 summary for configurations A--K in
Table~\ref{tab:ablations}. \textbf{(A)} Closure discrepancies and seed
variation for training and held-out cases. The $\Delta$ entries give the
mean $\pm$ sample standard deviation across ten seeds; Mean $s_\psi$ is the
sample standard deviation across seeds averaged over cases. Remarks
summarize the preceding closure curves. \textbf{(B)} RMS seed variation
$D_\psi$ over the indicated Reynolds number intervals, and the mean and
sample standard deviation $s$ of the best training loss. Each $D_\psi$ uses
a separate uniform grid of $20{,}001$ points. The selected 18-parameter
configuration J is bold.}
\label{tab:ladder-summary}
\end{table}

Table~\ref{tab:ladder-summary} confirms the closure curves. Over the training and held-out
range $13.1\leq\Relam\leq1498.7$, the selected closure (configuration J) reduces $D_\psi$ from
0.0708 for the unconstrained closure to 0.0032, about 95\%. The model is therefore selected by reproducibility rather than by training loss, since the unnormalized case reaches the
lowest loss (Section~\ref{subsec:results-fixed-coordinates}).

\subsection{Sensitivity to the supplied datasets}
\label{subsec:results-dataset-sensitivity}

The selected construction is applied without modification to Datasets~2
and~3; only the trajectories change. The endpoint values remain exact and
$\psi>1$ holds for every seed.

\begin{figure}[htbp]
\centering
\includegraphics[width=\linewidth]{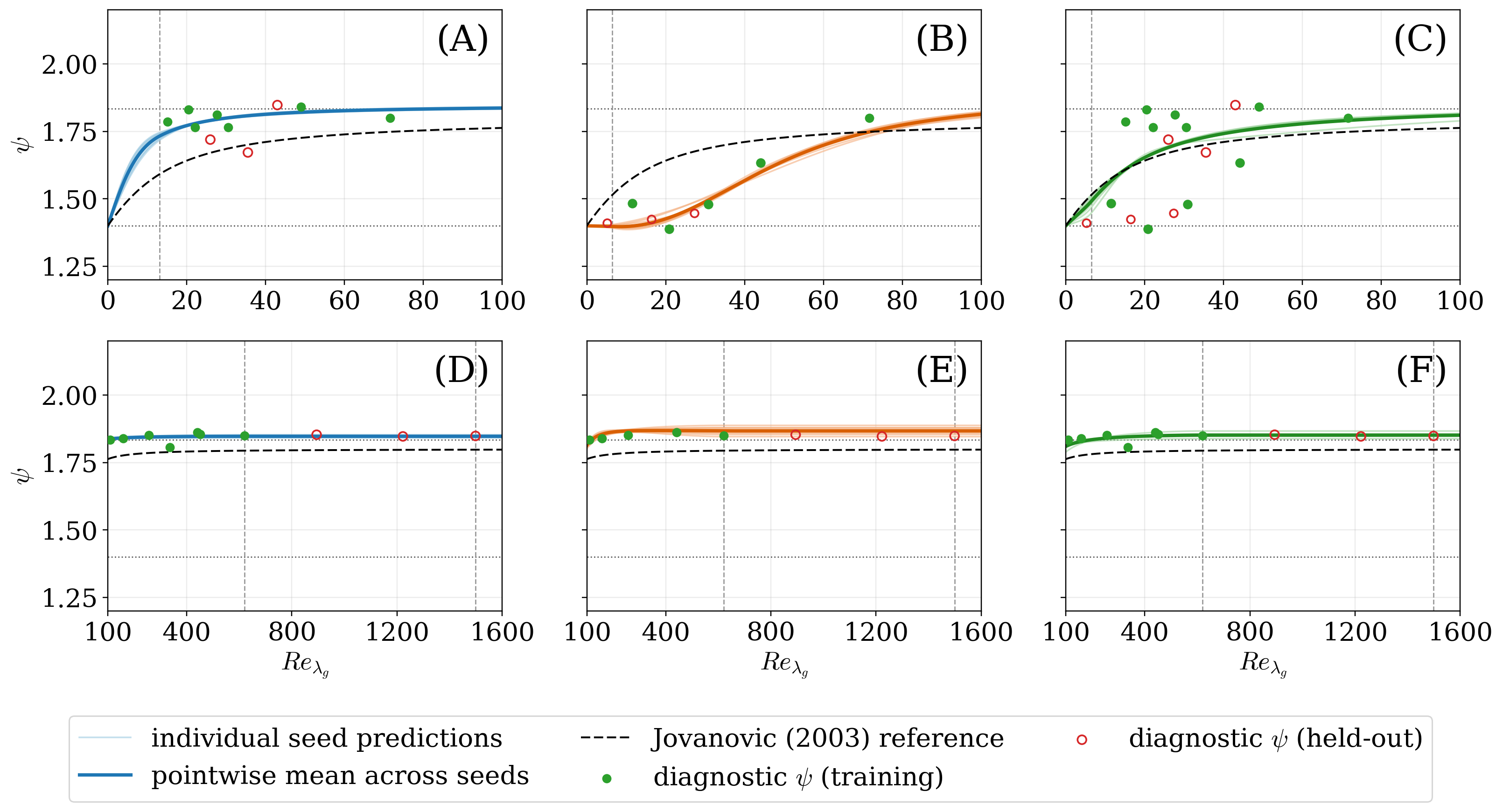}
\caption{Closures learned from the three datasets. Panels A and
D show Dataset~1, panels B and E show Dataset~2, and panels C and F show
Dataset~3. Panel ranges as in Fig.~\ref{fig:psi-capacity}.}
\label{fig:psi-dataset-collections}
\end{figure}

\begin{table}[htbp]
\centering
\resizebox{\linewidth}{!}{%
\begin{tabular}{lrrr rrr}
\toprule
& & & & \multicolumn{3}{c}{$D_\psi$} \\
\cmidrule(lr){5-7}
Dataset & Train/held out & Training $\Relam$ support & Mean best loss
& Below & Supported & Tail \\
\midrule
1 & 14/6 & 13.1--620 & $(4.67\pm0.82)\times10^{-5}$
& 0.0150 & 0.0030 & 0.0020 \\
2 & 10/6 & 6.44--620 & $(7.81\pm1.37)\times10^{-5}$
& 0.0020 & 0.0088 & 0.0077 \\
3 & 18/9 & 6.44--620 & $(4.66\pm0.13)\times10^{-4}$
& 0.0145 & 0.0063 & 0.0042 \\
\bottomrule
\end{tabular}}
\caption{Seed reproducibility of the selected framework for each dataset. The supported interval extends from the lowest Reynolds number
reached by the training trajectories to the largest initial Reynolds
number,
which enters the physics collocation at $t^*=0$. ``Below'' and ``Tail'' denote
the regions below and above that interval, respectively. $D_\psi$ is computed as in Table~\ref{tab:ladder-summary}B.}
\label{tab:dataset-sensitivity}
\end{table}

For each dataset, the ten seeds recover nearly the same closure, and the three datasets give different closures (Fig.~\ref{fig:psi-dataset-collections}, Table~\ref{tab:dataset-sensitivity}). The differences are largest below
$\Relam\approx24$, where the means differ by up to 0.35: the DNS-based
Dataset~2 gives a lower closure with a later rise, and the combined Dataset~3
lies in between. By $\Relam=620$, all three means approach 1.85. At low
Reynolds number, the DNS closure lies below several exponent-based diagnostic values; these assume a single constant closure value per case, whereas the forward
integration in Section~\ref{subsec:results-forward} tests the closure as a function of $\Relam$.

\subsection{\texorpdfstring{Forward evaluation of the learned closures}{Forward evaluation of the learned closures}}
\label{subsec:results-forward}

The closures are integrated forward without the state module
(Section~\ref{subsec:closure-extraction}), which tests them as standalone
closures. Dataset~3 contains the held-out
cases of Datasets~1 and~2, so all three closures are evaluated on its nine
held-out cases. We refer to the Dataset~3 closure as the combined closure.

\begin{figure}[htbp]
\centering
\includegraphics[width=\linewidth]{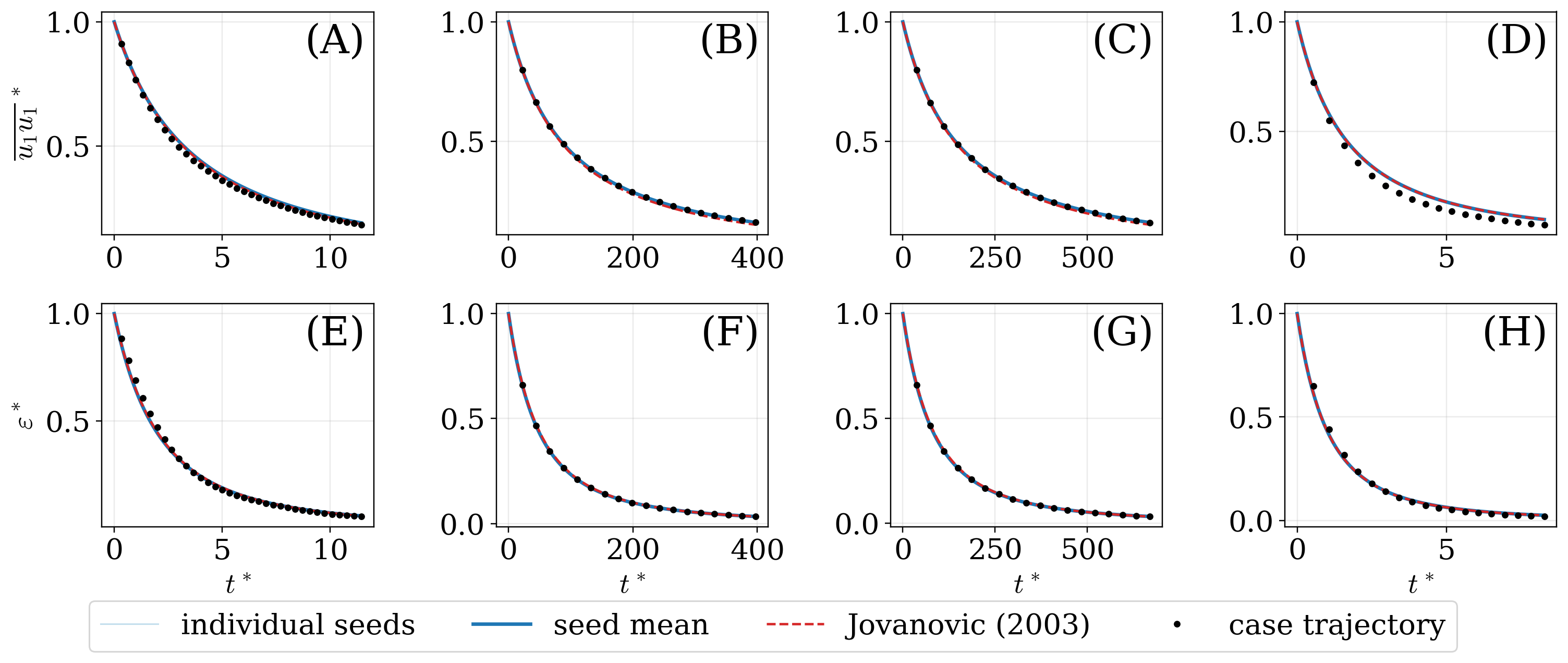}
\caption{Independent forward integration of the combined closure for four
randomly selected held-out cases. Panels A--D show $\ustar$ and panels E--H
show $\estar$. The columns correspond to wind Case~9, VDTT $1.7\times10^6$,
VDTT $4.8\times10^6$, and d128\_0.007.}
\label{fig:forward-combined}
\end{figure}

\begin{table}[htbp]
\centering
\resizebox{\linewidth}{!}{%
\begin{tabular}{@{}llrrrr@{}}
\toprule
 & & \multicolumn{4}{c}{Mean squared error} \\
\cmidrule(lr){3-6}
Held-out cases & State & Dataset~1 & Dataset~2 & Dataset~3 & Jovanovic \\
\midrule
VDTT (3) & $\ustar$ & $4.7\times10^{-7}$ & $1.0\times10^{-5}$ & $1.2\times10^{-6}$ & $6.2\times10^{-5}$ \\
 & $\estar$ & $7.1\times10^{-8}$ & $1.6\times10^{-6}$ & $1.9\times10^{-7}$ & $9.5\times10^{-6}$ \\
\midrule
Other cases (3) & $\ustar$ & $2.6\times10^{-4}$ & $1.7\times10^{-3}$\textsuperscript{$\dagger$} & $1.7\times10^{-4}$ & $2.2\times10^{-4}$ \\
 & $\estar$ & $1.5\times10^{-4}$ & $3.0\times10^{-4}$\textsuperscript{$\dagger$} & $1.1\times10^{-4}$ & $1.1\times10^{-4}$ \\
\midrule
DNS (3) & $\ustar$ & $2.6\times10^{-3}$\textsuperscript{$\dagger$} & $5.4\times10^{-5}$ & $1.1\times10^{-3}$ & $1.0\times10^{-3}$ \\
 & $\estar$ & $3.8\times10^{-4}$\textsuperscript{$\dagger$} & $5.9\times10^{-5}$ & $2.1\times10^{-4}$ & $2.0\times10^{-4}$ \\
\midrule
All held out (9) & $\ustar$ & $9.5\times10^{-4}$ & $5.8\times10^{-4}$ & $4.2\times10^{-4}$ & $4.4\times10^{-4}$ \\
 & $\estar$ & $1.8\times10^{-4}$ & $1.2\times10^{-4}$ & $1.1\times10^{-4}$ & $1.0\times10^{-4}$ \\
\bottomrule
\end{tabular}}
\caption{Mean squared \rev{errors in} $\ustar$ and $\estar$ \rev{from forward
integration on held-out cases. Errors are averaged across the cases in each
group. For Datasets~1--3, the table reports the mean of these group averages
from ten training runs.} Numbers in parentheses give the number of cases.
\rev{``Other cases'' includes} wind Case~9, the held-out Burattini case, and
Lavoie Rd35--WSA.
\textsuperscript{$\dagger$}Closure learned without data from this source.}
\label{tab:forward-errors}
\end{table}

Datasets~1 and~2 contain different low Reynolds number data, which
require different closure values at the same $\Relam$
(Fig.~\ref{fig:psi-dataset-collections}). The forward errors follow this
difference (Table~\ref{tab:forward-errors}): on the held-out DNS cases, the
Dataset~2 closure reduces the mean squared error of $\ustar$ from
$2.6\times10^{-3}$ to $5.4\times10^{-5}$ relative to the Dataset~1 closure,
and on the other held-out cases, the Dataset~1 closure reduces it from
$1.7\times10^{-3}$ to $2.6\times10^{-4}$ relative to the Dataset~2 closure.
Each closure is therefore accurate where its training data apply. The combined closure is trained on both and averages the conflicting values: at
$\Relam=20$, its mean is 1.65, between 1.77 for Dataset~1 and 1.43 for
Dataset~2. Averaged over all nine held-out cases, it gives the lowest total
error of the three closures, $4.2\times10^{-4}$ for $\ustar$ and
$1.1\times10^{-4}$ for $\estar$ (Fig.~\ref{fig:forward-combined}), although
within the DNS and VDTT groups, the Dataset~2 and Dataset~1 closures,
respectively, are more accurate. The \rev{model of} Jovanovi\'{c} et al. gives
errors
\rev{similar to those of} the combined closure. Refining the step to $\Delta\tstar=0.025$
changes the integrated states by less than 0.5\% of their initial values.
Finally, we assess the cost of the deployed closure network.

\subsection{Native closure cost}
\label{subsec:results-timing}

Because the seeds agree closely, the combined closure of seed 42 is written as a
scalar function in the C programming language and compared with the
\rev{closure of} Jovanovi\'{c} et al.
in the same source file and build. It evaluates the closure network, consisting
of coordinate, closure,
BC, and realizability modules, without PyTorch and matches the PyTorch closure
to within $2.89\times10^{-7}$ over $0\leq\Relam\leq10^6$.
Table~\ref{tab:native-cost} reports the timings on an Intel Core i7-1355U
using a 64-bit MSVC build with \texttt{/O2}, after an untimed warm-up.

\begin{table}[htbp]
\centering
\setlength{\tabcolsep}{2pt}
\begin{tabular}{lrrr}
\toprule
Implementation & \shortstack{Single\\evaluation (ns)}
& \shortstack{RK4 traj.\\($\mu$s)}
& \shortstack{10,000 traj.\\10 workers (ms)} \\
\midrule
Jovanovi\'{c} et al. & 1.21 & 8.42 & 18.19 \\
Combined C closure & 72.4 & 34.10 & 84.79 \\
\bottomrule
\end{tabular}
\caption{Native evaluation cost after untimed warm-up. \rev{To obtain the cost
of a single evaluation, we subtract the time taken
by an otherwise identical loop of no-op calls}. The RK4 value is the average per
trajectory across six benchmark cases. The batch uses equal-length 118-step trajectories and ten persistent workers, excluding pool creation and
destruction. Values are medians of three process runs.}
\label{tab:native-cost}
\end{table}

One evaluation of the learned closure costs about 72 ns, compared with
about 1 ns for the \rev{expression of} Jovanovi\'{c} et al. Including the state
update, a complete RK4 trajectory takes about four times as
\rev{long with the learned closure} as with the \rev{closure of} Jovanovi\'{c}
et al. (34 versus 8.4~$\mu$s), and ten workers integrate 10,000 trajectories
in 85~ms.

\section{Discussion}
\label{sec:discussion}

\subsection{\texorpdfstring{Role of the modules}{Role of the modules}}
\label{subsec:discussion-constraints}

The BC and realizability modules impose the boundary values and
realizability by construction, independently of the data and of training. Moreover,
the interior of the closure is learned and depends on the input coordinates.
\rev{The learned closure depends on the scaling of the fixed input coordinates}
even when the training loss decreases. In \rev{decaying HIT}, \rev{the closure had
only one} scalar input \rev{and still showed} a hump \rev{that varied across seeds}.
Learned coordinates removed it and made the closure
reproducible across seeds at all tested capacities, which is why the
coordinate module is part of the framework.

\subsection{\texorpdfstring{Dependence on the data}{Dependence on the data}}
\label{subsec:discussion-normalization}

With the same framework, different datasets produce different learned
closures, particularly at low Reynolds number.
The observed differences suggest that the instantaneous Taylor scale
Reynolds number alone may not capture all differences between the
trajectories. One possible cause is the dependence of
decaying HIT on its initial conditions. Because the modules are
separate, the closure module can take further inputs, such as a parameter of
the initial spectrum, to test this hypothesis.

\subsection{\texorpdfstring{Limitations and extension}{Limitations and extension}}

Reproducibility across seeds alone
does not show that the learned closure is correct\rev{. We therefore tested
the learned closures} on held-out trajectories \rev{using an independent RK4
solver} without the state module.
The learned closure network is more costly to evaluate than an
algebraic expression. To reduce this cost, symbolic
regression may provide an algebraic approximation
for the closure network \citep{unal2026symbolic}.
Finally, the chamber, Burattini,
Lavoie, Comte-Bellot--Corrsin, and Sinhuber cases are reconstructed from
power-law decay exponents, so each implies a constant closure value over
its trajectory.

\rev{The ablation studies in decaying HIT showed the importance of
individual modules. In homogeneous anisotropic turbulence, new modules
can enforce bounds on Reynolds stress anisotropy and other physical constraints \citep{lumley1978}.}
\newpage
\section{Conclusions}
\label{sec:conclusion}

We presented a \rev{method} for constructing a deployable hard-constrained
\rev{closure for the dissipation rate transport equation}.
\rev{We learned the closure function $\psi$ from state trajectories of
decaying HIT without using reference values of $\psi$ as training targets.
After training, we
exported the closure as a standalone C function of $\Relam$.}

The prescribed boundary values held exactly, and growth of the Taylor
microscale was guaranteed by $\psi>1$ for every seed.
\rev{With fixed input coordinates, the closure still developed a hump
whose size varied across seeds. Increasing the network width made the hump
more pronounced. Learning the coordinates removed the hump and gave nearly
the same closure across seeds at every width tested.}

\rev{The closures learned from different datasets differed most at low
Reynolds numbers. In independent forward integration, the closure trained
on the combined data gave the lowest mean errors of the three learned
closures across all held-out cases. Since the DNS-based closure was more accurate
on the held-out DNS cases, we can add inputs to the closure module to test
the dependency of $\psi$ on initial conditions. We aim to extend this framework to 
model the closures of unclosed terms in the transport equations of Reynolds
stresses and turbulent dissipation rate for more complex turbulent flows by
enforcing additional physical constraints, such as Lumley realizability.}

\section*{Acknowledgement}
This study was supported by the Scientific and Technological Research Council of Türkiye (T\"UB\.{I}TAK), Project No. 124M535, and by the EU co-funded partnership CONNECT-NM under Grant Agreement No. 101165375. Tamerlan Hajiyev thanks {\"O}zye{\u{g}}in University for providing his scholarship. We thank Veli Can Coşar and Yeabsira Simeka Maru for providing HIT chamber data.

\section*{Credit authorship contribution statement}
\textbf{Tamerlan Hajiyev:} Writing – original draft, Visualization, Validation, Software, Methodology, Investigation, Formal analysis, Data curation, Conceptualization; \textbf{Emre Sefer:} Supervision, Software, Methodology, Writing – review \& editing;  \textbf{Ivan Otić:} Writing – review \& editing, Methodology, Formal analysis; \textbf{Ismail Ari:} Writing – review \& editing, Methodology, Conceptualization; \textbf{Özgür Ertunç:} Writing – review \& editing, Supervision, Project administration, Methodology, Funding acquisition, Formal analysis, Data curation, Conceptualization.

\section*{Declaration of competing interest}
We have no conflicts of interest to declare.

\section*{Data availability}
Training and test data \rev{ are available at \doi{10.5281/zenodo.24440915} and the trained closure model is available at \doi{10.5281/zenodo.24461958} on Zenodo.}

%\section*{Acknowledgments}
%[To be completed.]



\bibliographystyle{elsarticle-harv}
\bibliography{refs}

@article{Speziale_Sarkar_Gatski_1991, title={Modelling the pressure–strain correlation of turbulence: an invariant dynamical systems approach}, volume={227}, DOI={10.1017/S0022112091000101}, journal={Journal of Fluid Mechanics}, author={Speziale, Charles G. and Sarkar, Sutanu and Gatski, Thomas B.}, year={1991}, pages={245–272}}

@article{Hanjalic_Launder_1976, title={Contribution towards a Reynolds-stress closure for low-Reynolds-number turbulence}, volume={74}, DOI={10.1017/S0022112076001961}, number={4}, journal={Journal of Fluid Mechanics}, author={Hanjalić, K. and Launder, B. E.}, year={1976}, pages={593–610}}

@article{LAUNDER1989,
title = {Second-moment closure: present… and future?},
journal = {International Journal of Heat and Fluid Flow},
volume = {10},
number = {4},
pages = {282-300},
year = {1989},
issn = {0142-727X},
doi = {10.1016/0142-727X(89)90017-9},
author = {Brian E. Launder}}

@article{mortari2017tfc,
  author  = {Mortari, Daniele},
  title   = {The theory of connections: connecting points},
  journal = {Mathematics},
  volume  = {5},
  number  = {4},
  pages   = {57},
  year    = {2017},
  doi     = {10.3390/math5040057}
}

@article{leake2020deeptfc,
  author  = {Leake, Carl and Mortari, Daniele},
  title   = {Deep theory of functional connections: a new method for estimating the solutions of partial differential equations},
  journal = {Machine Learning and Knowledge Extraction},
  volume  = {2},
  number  = {1},
  pages   = {37--55},
  year    = {2020},
  doi     = {10.3390/make2010004}
}

@article{jovanovic2003,
  author  = {Jovanovic, J. and Oti{\'c}, I. and Bradshaw, P.},
  title   = {On the Anisotropy of Axisymmetric Strained Turbulence in the Dissipation Range},
  journal = {Journal of Fluids Engineering},
  volume  = {125},
  number  = {3},
  pages   = {401--413},
  year    = {2003},
  doi     = {10.1115/1.1568355}
}

@article{lu2021deeponet,
  author  = {Lu, Lu and Jin, Pengzhan and Pang, Guofei and Zhang, Zhongqiang and Karniadakis, George Em},
  title   = {Learning nonlinear operators via {DeepONet} based on the universal approximation theorem of operators},
  journal = {Nature Machine Intelligence},
  volume  = {3},
  pages   = {218--229},
  year    = {2021},
  doi     = {10.1038/s42256-021-00302-5}
}

@article{taassob2024,
  author  = {Taassob, Arsalan and Kumar, Anuj and Gitushi, Kevin M. and Ranade, Rishikesh and Echekki, Tarek},
  title   = {A {PINN}--{DeepONet} framework for extracting turbulent combustion closure from multiscalar measurements},
  journal = {Computer Methods in Applied Mechanics and Engineering},
  volume  = {429},
  pages   = {117163},
  year    = {2024},
  doi     = {10.1016/j.cma.2024.117163}
}

@article{gunseren,
  author  = {G{\"u}nseren, Deniz and Ertun{\c{c}}, {\"O}zg{\"u}r and Ari, Ismail and Atik, Atakan Ataman and Muhtaroglu, Nitel and Oti{\'c}, Ivan},
  title   = {Constructing a {Reynolds} stress model of decaying homogeneous isotropic turbulence using physics informed neural network},
  journal = {Journal of Computational Physics},
  volume  = {553},
  pages   = {114725},
  year    = {2026},
  doi     = {10.1016/j.jcp.2026.114725}
}

@book{pope2000,
  author    = {Pope, Stephen B.},
  title     = {Turbulent Flows},
  publisher = {Cambridge University Press},
  address   = {Cambridge},
  year      = {2000},
  doi       = {10.1017/CBO9780511840531}
}

@article{raissi2019pinn,
  author  = {Raissi, M. and Perdikaris, P. and Karniadakis, G. E.},
  title   = {Physics-informed neural networks: A deep learning framework for solving forward and inverse problems involving nonlinear partial differential equations},
  journal = {Journal of Computational Physics},
  volume  = {378},
  pages   = {686--707},
  year    = {2019},
  doi     = {10.1016/j.jcp.2018.10.045}
}

@article{karniadakis2021,
  author  = {Karniadakis, George Em and Kevrekidis, Ioannis G. and Lu, Lu and Perdikaris, Paris and Wang, Sifan and Yang, Liu},
  title   = {Physics-informed machine learning},
  journal = {Nature Reviews Physics},
  volume  = {3},
  number  = {6},
  pages   = {422--440},
  year    = {2021},
  doi     = {10.1038/s42254-021-00314-5}
}

@article{lagaris1998,
  author  = {Lagaris, I. E. and Likas, A. and Fotiadis, D. I.},
  title   = {Artificial neural networks for solving ordinary and partial differential equations},
  journal = {IEEE Transactions on Neural Networks},
  volume  = {9},
  number  = {5},
  pages   = {987--1000},
  year    = {1998},
  doi     = {10.1109/72.712178}
}

@article{vangastelen2024,
  author  = {van Gastelen, Toby and Edeling, Wouter and Sanderse, Benjamin},
  title   = {Energy-conserving neural network for turbulence closure modeling},
  journal = {Journal of Computational Physics},
  volume  = {508},
  pages   = {113003},
  year    = {2024},
  doi     = {10.1016/j.jcp.2024.113003}
}

@article{saffman1967,
  author  = {Saffman, P. G.},
  title   = {Note on Decay of Homogeneous Turbulence},
  journal = {Physics of Fluids},
  volume  = {10},
  number  = {6},
  pages   = {1349},
  year    = {1967},
  doi     = {10.1063/1.1762284}
}

@article{batchelor1948,
  author  = {Batchelor, G. K. and Townsend, A. A.},
  title   = {Decay of turbulence in the final period},
  journal = {Proceedings of the Royal Society of London. Series A. Mathematical and Physical Sciences},
  volume  = {194},
  number  = {1039},
  pages   = {527--543},
  year    = {1948},
  doi     = {10.1098/rspa.1948.0095}
}

@article{burattini2006,
  author  = {Burattini, Paolo and Lavoie, Philippe and Agrawal, Amit and Djenidi, Lyazid and Antonia, Robert A.},
  title   = {Power law of decaying homogeneous isotropic turbulence at low {R}eynolds number},
  journal = {Physical Review E},
  volume  = {73},
  number  = {6},
  pages   = {066304},
  year    = {2006},
  doi     = {10.1103/PhysRevE.73.066304}
}

@article{lavoie2007,
  author  = {Lavoie, Philippe and Djenidi, Lyazid and Antonia, Robert A.},
  title   = {Effects of initial conditions in decaying turbulence generated by passive grids},
  journal = {Journal of Fluid Mechanics},
  volume  = {585},
  pages   = {395--420},
  year    = {2007},
  doi     = {10.1017/S0022112007006763}
}

@article{cbc1971,
  author  = {Comte-Bellot, Genevi{\`e}ve and Corrsin, Stanley},
  title   = {Simple {E}ulerian time correlation of full- and narrow-band velocity signals in grid-generated, `isotropic' turbulence},
  journal = {Journal of Fluid Mechanics},
  volume  = {48},
  number  = {2},
  pages   = {273--337},
  year    = {1971},
  doi     = {10.1017/S0022112071001599}
}

@article{sinhuber2015,
  author  = {Sinhuber, Michael and Bodenschatz, Eberhard and Bewley, Gregory P.},
  title   = {Decay of turbulence at high {R}eynolds numbers},
  journal = {Physical Review Letters},
  volume  = {114},
  number  = {3},
  pages   = {034501},
  year    = {2015},
  doi     = {10.1103/PhysRevLett.114.034501}
}

@article{yoffe2018,
  author  = {Yoffe, Samuel R. and McComb, W. David},
  title   = {Onset criteria for freely decaying isotropic turbulence},
  journal = {Physical Review Fluids},
  volume  = {3},
  number  = {10},
  pages   = {104605},
  year    = {2018},
  doi     = {10.1103/PhysRevFluids.3.104605}
}

@article{ling2016,
  author  = {Ling, Julia and Kurzawski, Andrew and Templeton, Jeremy},
  title   = {Reynolds averaged turbulence modelling using deep neural networks with embedded invariance},
  journal = {Journal of Fluid Mechanics},
  volume  = {807},
  pages   = {155--166},
  year    = {2016},
  doi     = {10.1017/jfm.2016.615}
}

@inproceedings{durkan2019neural,
  author    = {Durkan, Conor and Bekasov, Artur and Murray, Iain and Papamakarios, George},
  title     = {Neural Spline Flows},
  booktitle = {Advances in Neural Information Processing Systems},
  volume    = {32},
  pages     = {7511--7522},
  year      = {2019},
  url       = {https://proceedings.neurips.cc/paper/2019/hash/7ac71d433f282034e088473244df8c02-Abstract.html}
}

@inproceedings{ramasinghe2022bernstein,
  author    = {Ramasinghe, Sameera and Fernando, Kasun and Khan, Salman and Barnes, Nick},
  title     = {Robust Normalizing Flows Using {Bernstein}-Type Polynomials},
  booktitle = {Proceedings of the 33rd British Machine Vision Conference},
  publisher = {BMVA Press},
  pages     = {532},
  year      = {2022},
  note      = {arXiv:2102.03509},
  url       = {https://pdf-files.bmvc2022.org/0532.pdf}
}

@article{farouki1996stability,
  author  = {Farouki, Rida T. and Goodman, T. N. T.},
  title   = {On the Optimal Stability of the {Bernstein} Basis},
  journal = {Mathematics of Computation},
  volume  = {65},
  number  = {216},
  pages   = {1553--1566},
  year    = {1996},
  doi     = {10.1090/S0025-5718-96-00759-4}
}

@article{unal2026symbolic,
  author  = {{\"U}nal, Yunus Emre and Ertun{\c{c}}, {\"O}zg{\"u}r and Ari, Ismail and Oti{\'c}, Ivan},
  title   = {Order of Magnitude Analysis and Data-Based Physics-Informed Symbolic Regression for Turbulent Pipe Flow},
  journal = {arXiv preprint arXiv:2602.17082},
  year    = {2026},
  doi     = {10.48550/arXiv.2602.17082}
}

@article{zhang2026bluff,
  author  = {Zhang, Zhen and K{\"a}ufer, Theo and Ronglan, Louise and Triantafyllou, Michael S. and Karniadakis, George Em},
  title   = {Generalizable turbulence closures across bluff-body shapes by {PINN}-based solver-agnostic training},
  journal = {arXiv preprint arXiv:2607.04491},
  year    = {2026},
  doi     = {10.48550/arXiv.2607.04491}
}

@article{mansour1994,
  author  = {Mansour, Nagi N. and Wray, Alan A.},
  title   = {Decay of isotropic turbulence at low {R}eynolds number},
  journal = {Physics of Fluids},
  volume  = {6},
  number  = {2},
  pages   = {808--814},
  year    = {1994},
  doi     = {10.1063/1.868319}
}

@article{mcconkey2025realisability,
  author  = {McConkey, Ryley and Kalia, Nikhila and Yee, Eugene and Lien, Fue-Sang},
  title   = {Realisability-informed machine learning for turbulence anisotropy mappings},
  journal = {Journal of Fluid Mechanics},
  volume  = {1019},
  pages   = {A49},
  year    = {2025},
  doi     = {10.1017/jfm.2025.10618}
}

@inproceedings{cosar2026ilass,
  author    = {Co{\c{s}}ar, Veli Can and Simeka, Yeabsira and Ertun{\c{c}}, {\"O}zg{\"u}r},
  title     = {Correlation-Based Turbulent {Weber} Number and {Taylor} {Reynolds} Number in a Fan-Stirred Isotropic Chamber: {Hinze}'s Relation and Breakup Regimes},
  booktitle = {Proceedings of ILASS Europe 2026, 34th Conference on Liquid Atomization and Spray Systems},
  address   = {Lisbon, Portugal},
  pages     = {1--8},
  year      = {2026}
}

@phdthesis{ertunc2007,
  author  = {Ertun{\c{c}}, {\"O}zg{\"u}r},
  title   = {Experimental and Numerical Investigations of Axisymmetric Turbulence},
  school  = {Universit{\"a}t Erlangen-N{\"u}rnberg},
  address = {Erlangen, Germany},
  year    = {2007}
}

@article{launder1975,
  author  = {Launder, B. E. and Reece, G. J. and Rodi, W.},
  title   = {Progress in the development of a Reynolds-stress turbulence closure},
  journal = {Journal of Fluid Mechanics},
  volume  = {68},
  number  = {3},
  pages   = {537--566},
  year    = {1975},
  doi     = {10.1017/S0022112075001814}
}

@article{pope1975,
  author  = {Pope, S. B.},
  title   = {A more general effective-viscosity hypothesis},
  journal = {Journal of Fluid Mechanics},
  volume  = {72},
  number  = {2},
  pages   = {331--340},
  year    = {1975},
  doi     = {10.1017/S0022112075003382}
}

@article{lumley1978,
  author  = {Lumley, J. L.},
  title   = {Computational Modeling of Turbulent Flows},
  journal = {Advances in Applied Mechanics},
  volume  = {18},
  pages   = {123--176},
  year    = {1978},
  doi     = {10.1016/S0065-2156(08)70266-7}
}

@article{wang2021pideeponet,
  author  = {Wang, Sifan and Wang, Hanwen and Perdikaris, Paris},
  title   = {Learning the solution operator of parametric partial differential equations with physics-informed {DeepONets}},
  journal = {Science Advances},
  volume  = {7},
  number  = {40},
  pages   = {eabi8605},
  year    = {2021},
  doi     = {10.1126/sciadv.abi8605}
}

\end{document}